\documentclass{article}
\usepackage{arxiv}
\usepackage[T1]{fontenc}    
\usepackage{setspace}
\usepackage{hyperref}       
\usepackage{url}            
\usepackage{booktabs}       
\usepackage{amsfonts}       
\usepackage{nicefrac}       
\usepackage{microtype}      
\usepackage{lipsum}		
\usepackage{graphicx}
\usepackage{natbib}
\usepackage{dsfont}
\usepackage{doi}
\usepackage{caption}
\usepackage{amsthm,amsmath,amssymb}
\usepackage{graphicx,bm}
\usepackage{color}
\usepackage{lscape}
\usepackage{rotating}
\usepackage{setspace}
\usepackage{algorithm}
\usepackage{algpseudocode}

\title{Logistic Multidimensional Data Analysis for Ordinal Response Variables using a Cumulative Link function}

\author{ \href{https://orcid.org/0000-0001-7308-6210}{\includegraphics[scale=0.06]{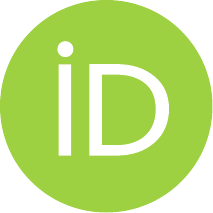}\hspace{1mm}Mark de Rooij}\\
	Methodology and Statistics department\\
	Leiden University\\
	The Netherlands \\
	\texttt{rooijm@fsw.leidenuniv.nl} \\
	\And
	\href{https://orcid.org/0000-xxxx-xxxx-xxxx}{\includegraphics[scale=0.06]{orcid.pdf}\hspace{1mm}Ligaya Breemer} \\
	Methodology and Statistics department\\
	Leiden University\\
	The Netherlands \\
	\texttt{l.breemer@fsw.leidenuniv.nl} \\	
	\And
	\href{https://orcid.org/0000-xxxx-xxxx-xxxx}{\includegraphics[scale=0.06]{orcid.pdf}\hspace{1mm}Dion Woestenburg} \\
	Methodology and Statistics department\\
	Leiden University\\
	The Netherlands \\
	\texttt{dionwoestenburg@hotmail.com} \\	
	\And
	\href{https://orcid.org/0000-0002-8062-538x}{\includegraphics[scale=0.06]{orcid.pdf}\hspace{1mm}Frank Busing} \\
	Methodology and Statistics department\\
	Leiden University\\
	The Netherlands \\
	\texttt{busing@fsw.leidenuniv.nl} \\	
}

\renewcommand{\headeright}{}
\renewcommand{\undertitle}{}
\renewcommand{\shorttitle}{Cumulative Logistic Multidimensional Data Analysis}

\makeatletter
\input{size12.clo}
\makeatother

\begin{document}
\maketitle

\begin{abstract}
We present a multidimensional data analysis framework for the analysis of ordinal response variables. Underlying the ordinal variables, we assume a continuous latent variable, leading to cumulative logit models. The framework includes unsupervised methods, when no predictor variables are available, and supervised methods, when predictor variables are available. We distinguish between dominance variables and proximity variables, where dominance variables are analyzed using inner product models, whereas the proximity variables are analyzed using distance models. An expectation-majorization-minimization algorithm is derived for estimation of the parameters of the models. We illustrate our methodology with three empirical data sets highlighting the advantages of the proposed framework. A simulation study is conducted to evaluate the performance of the algorithm. 
\end{abstract}


\newpage

\section{Introduction}

In many fields of study, ordered categorical variables, also called ordinal variables, are collected. In medicine, for example, patients can be  classified as, say, severely, moderately, or mildly ill \citep{anderson1981regression}. In the social and behavioural sciences, commonly Likert scales are used that have response categories such as "strongly disagree" (SD), "disagree" (D), "neutral" (N), "agree" (A), and "strongly agree" (SA). There is an ordering between these categories, but differences between these categories are unknown. It is standard practice to give numerical codes to the categories, such as 1, 2, 3, 4, 5, and subsequently perform a standard numerical analysis. In the context of regression modelling, \cite{liddell2018analyzing} argue that the analysis of ordinal response variables through linear models can lead to distorted effect sizes, inflated Type-I errors, and inversions of differences between groups. 

\begin{figure}[h]
\begin{center}
\includegraphics[width = .5\textwidth]{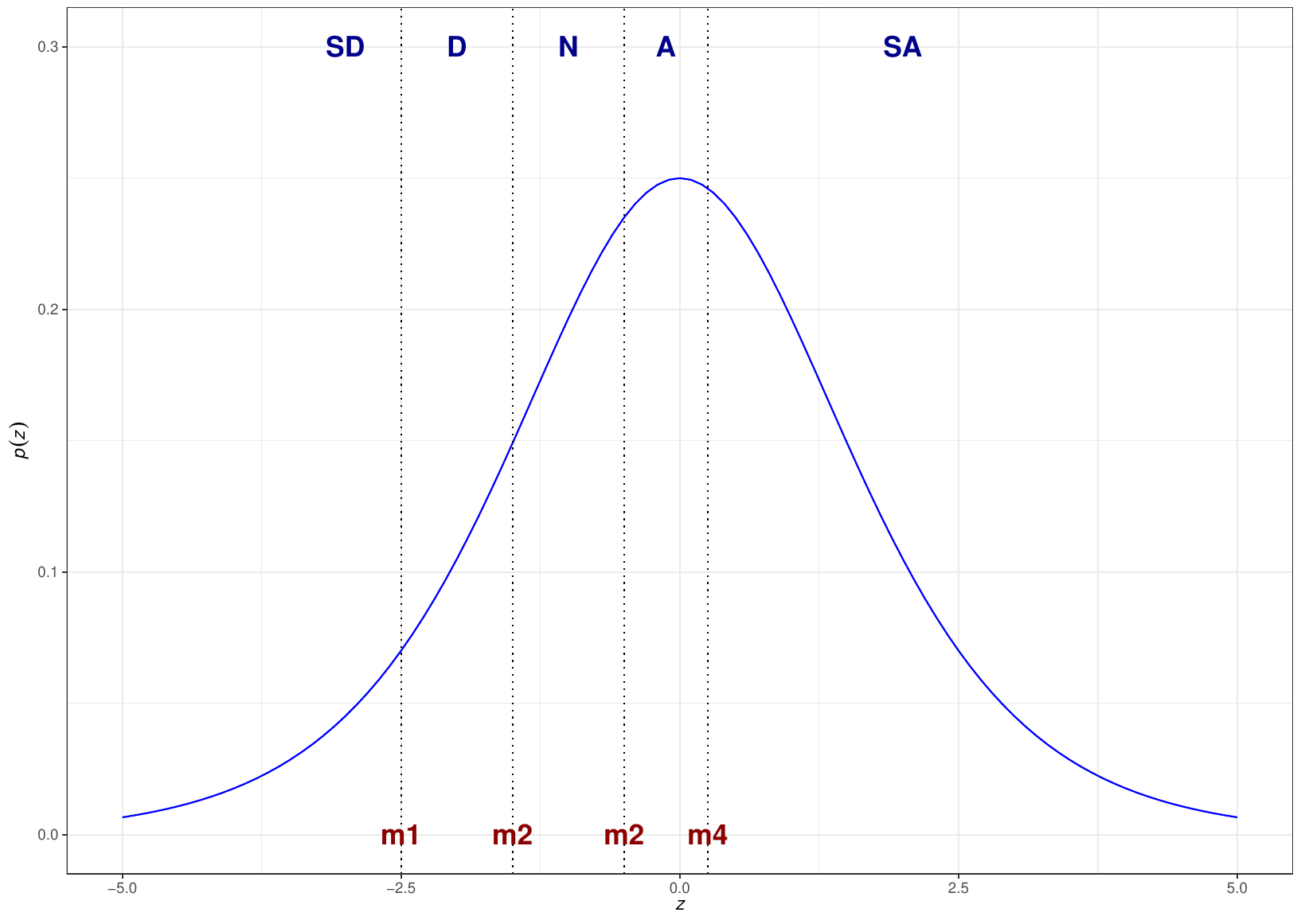}
\caption{Probability density function for a continuous latent variable $z$ with thresholds (indicated by the vertical lines) giving rise to an observed ordered categorical variable with categories, strongly disagree (SD), disagree (D), neutral (N), agree (A), and strongly agree (SA).}
\label{fig:latvar}
\end{center}
\end{figure}

Underlying many ordinal variables, a continuous variable can be assumed. This is a \emph{latent variable}, as we only observe the ordinal scores not the numerical ones. In Figure \ref{fig:latvar}, we show the density of such a latent numerical variable. Instead of the numerical values, we observe categories such as SD, D, N, A, and SA. The continuous underlying variable is partitioned through a set of cut-points or \emph{thresholds} into a set of categories. In Figure \ref{fig:latvar}, the thresholds are shown as vertical dashed lines. All responses falling between two thresholds invoke the same response category. More formally, let $z$ be the continuous latent variable. Define a set of thresholds $-\infty = m_0 < m_1<  \ldots < m_C = \infty$ such that an observed ordinal response $y$ satisfies
\[
y = c \ \  \mathrm{if} \ m_{c-1} \leq z < m_c
\]
for $c = 1, \ldots, C$. 

In regression modeling of an ordinal response variable, the following model for the latent variable is assumed \citep{anderson1981regression}
\[
z = \bm{x}'\bm{\beta} + \epsilon,
\]
where $\epsilon$ is an independent and identically distributed error term with cumulative density function $F$.  This regression model for the latent response variable implies
\[
P(y = c) = P(m_{c-1} < z \leq m_c) = F\left(m_c -  \bm{x}'\bm{\beta}  \right) - F\left(m_{c-1} -  \bm{x}'\bm{\beta}  \right). 
\]
It follows that 
\[
F^{-1}\left(P(y \leq c|\bm{x}) \right) = m_c -  \bm{x}'\bm{\beta}, 
\]
where $\bm{x}'\bm{\beta}$ is the \emph{structural part} of the model. 

In regression modelling, the \emph{de facto} default choice for the analysis of categorical response variables are logistic models \citep{agresti2002categorical}. For binary response variables, standard binary logistic regression models have been developed and these have been extended for ordinal variables and nominal variables \citep[see][chapter 7]{agresti2002categorical}. Logistic models have the advantage that detailed interpretation in terms of changes in log-odds is possible. Such an interpretation is not available for, for example, probit models that use the cumulative density of the normal distribution. Otherwise, the fit of logit and probit models is usually very similar \citep{agresti2002categorical}. In logistic regression models for ordinal variables, we use the cumulative density function of the logistic distribution, such that $F$ equals
\[
F(\eta) = \frac{1}{1 + \exp(-\eta)}
\]
and the corresponding regression model is known as the \emph{proportional odds model}, or, more generally, the cumulative logistic regression model \citep{walker1967estimation, mccullagh1980regression, anderson1981regression, agresti2002categorical}. 

In many investigations, \emph{multiple response variables} are collected. Researchers often analyze the response variables separately, but because the response variables are correlated this might not be an optimal strategy. Multidimensional data analysis refers to a set of data analysis techniques representing the multivariate data in a low dimensional, often Euclidean, space. The $R$ response variables are analyzed together and the results are represented in an $S$-dimensional space, where $S < R$. In the low dimensional representation, the associations (i.e., correlations) between response variables are modelled.  

For the analysis of data it is important to distinguish between two types of response processes \citep{coombs1964theory, polak2011}. In a unipolar or cumulative scale or map,  responses are monotonically related to the position of the person on the map. The response variables are so-called \emph{dominance variables}. Mathematical problems are a typical example of dominance items where subjects with a higher mathematical ability have a higher probability of solving the problem correctly. 
In a bipolar scale or map, the variable responses are characterized by the proximity between the variable and the respondent: The responses are single-peaked functions of the distance between the position of a variable and the position of a person.  The variables are so-called \emph{proximity variables}. For dominance variables the subjects are partitioned into homogeneous groups, that is, all subjects with a fixed response constitute a homogeneous group. For proximity items, reasons to answer totally disagree might differ between the respondents. Respondents who disagree therefore do not necessarily constitute a homogeneous group. 

In classical multivariate analysis, principal component analysis \citep[PCA, ][]{pearson1901principal, hotelling1936simplified, jolliffe2002principal} is the standard multidimensional data analysis tool for the analysis of dominance variables whereas multidimensional unfolding \citep[MDU, ][]{heiser1981unfolding, busing2010advances} is the standard tool for the analysis of proximity variables. 

In principal component analysis, a data set is summarized by reducing the dimensionality using a set of \emph{principal components}, that are, linear combinations of the original variables, that explain as much of the original variability as possible. The data are summarized by principal scores and variable loadings. The principal scores are scores of the individuals on the principal components, while the variable loadings give the relation between the principal components and the original response variables. PCA solutions can be graphically represented through so-called biplots \citep{gabriel1971biplot, gower1996biplots, gower2011understanding}, where the row objects (observations, participants, individuals) are represented as points in a Euclidean space and the columns (variables, items) as vectors or variable axes. Estimated values of the responses can be obtained from the biplot by orthogonal projection of the row object points onto the variable axes. 

Another multidimensional data analysis approach is multidimensional unfolding. MDU is targeted towards proximity variables. Where PCA uses an inner-product representation of the data, MDU uses a distance representation. MDU is a generalization of multidimensional scaling to two-way two-mode data matrices \citep{CarrollArabie1980}. In MDU, the data are approximated by the distance between so-called ideal points for the row objects (i.e., participants) and points for the response variables, that is, the distances between the two sets of points. MDU solutions can also be graphically represented by biplots, where both the row-objects and the variables are represented by points in a Euclidean space. Estimated values can be obtained by inspecting the distances between the two sets of points. 

When besides the response variables also predictor variables are available on the row objects, we can constrain the PCA or MDU to incorporate this information. The principal scores or the ideal points are restricted to be (linear) functions of the predictor variables. When we constrain PCA in such a manner, the resulting model is known as reduced rank regression \citep[RRR;][]{izenman1975reduced, tso1981reduced} or redundancy analysis \citep{wollenberg1977redundancy}. Reduced rank regression models can also be represented graphically, by so-called triplots \citep{braak1994biplots}. 
When we constrain MDU in such manner, we obtain a model known as restricted multidimensional unfolding \citep[RMDU;][]{busing2010restricted}. These restricted multidimensional unfolding models can also be graphically represented by triplots. 

PCA, MDU and its constrained versions, RRR and RMDU, are usually estimated by least squares methods. For categorical response variables, however, linear models estimated with least squares are not optimal and might lead to distorted effect sizes, type-I errors and inversion of effects \citep{liddell2018analyzing}. Logistic models estimated using maximum likelihood offer an alternative. 

For binary data, several authors \citep{schein2003generalized, deleeuw2006principal, landgraf2020dimensionality}  proposed PCA using the binomial negative log-likelihood as loss function. \cite{collins2001generalization} proposed a generalization of PCA to the exponential family to deal with, for example, binary data or integer-valued data such as count data. As far as we know, only \cite{vicente2014logistic} investigated an exponential family generalizations of PCA for ordinal response variables including a biplot visualization. 

For constrained PCA, that is reduced rank regression or redundancy analysis, a logistic model has been proposed for binary data by \cite{derooij2023new} and \citep{yee2003reduced} proposed a wider class of models for the exponential family, similar to generalized linear models \citep{McCullagh1989GLM}. As far as we know, there are no exponential family generalizations of these reduced rank models for ordinal response variables. 

For MDU, several attempts can be found in the literature to exponential family multidimensional unfolding models.  \cite{andrich1988application, takane1998choice, desarbo1986simple} defined MDU models for binary variables using squared distances. \cite{andrich1988application} proposed a unidimensional model that does not allow for predictor variables. \cite{takane1998choice} and \cite{desarbo1986simple} describe generalizations to multiple dimensions that can include predictors. \cite{derooijwoestenburgbusing2022} defined a model on the basis of (unsquared) distances for binary data, both with and without predictor variables. As far as we know, there are no exponential family generalizations of MDU for ordinal response variables.

In conclusion, there have been several attempts to define exponential family models for PCA and MDU and their constraint versions that include predictor variables. These attempts mainly focus on binary response variables, but some include also other types of variables \citep{collins2001generalization, yee2003reduced} like count variables. However, no exponential family generalizations exist for PCA or MDU of ordinal responses, neither with or without predictor variables. 

The goal of this paper is to fill this gap and propose multidimensional models for ordinal response variables in the exponential family. We will develop models for multivariate ordinal dominance variables (i.e., PCA and RRR) and proximity variables (i.e., MDU and RMDU). Both models without predictor variables (i.e., PCA and MDU) and with predictor variables (i.e., RRR and RMDU) will be presented. Along the algebraic formulation, we will also develop biplot methodology for visualization of the models. One unified algorithm for maximum likelihood estimation of model parameters will be developed and tested, where in the lower level iterations updates differ between the four approaches. To illustrate the multidimensional models and the difference between the dominance and proximity perspective, we will apply the models on several data sets. In the first application, we have cognitive data and use the dominance approach. The second application highlights the difference between the dominance and proximity approaches using behavioural response variables. The third application shows in detail the proximity approach where the response variables concern attitudes concerning the environment. Biplots for all three applications will be discussed in detail. 

The outline of this manuscript is as follows. In Section \ref{sec:clmda}, we will propose a family of geometric, multidimensional, models for multivariate ordinal data. We distinguish between dominance and proximity response variables and between models with and without predictor variables. Properties of the model will be derived. We briefly discuss model selection and discuss in detail the visualization of the models using biplots. In Section \ref{sec:estimation}, a unified algorithm for maximum likelihood estimation of model parameters will be presented. In Section \ref{sec:applications}, we show three applications. The first application considers data from students performance on a statistics exam. The second and third data set are from the International Social Survey Programme \citep{ISSP2020data}. We test our algorithm using simulated data in Section \ref{sec:simulations}. We end this paper with some discussion and conclusions.  

\section{Cumulative Logistic Multidimensional Models}\label{sec:clmda}

We consider a set of ordinal variables with observed values $y_{ir}$ ($i = 1,\ldots, N$, $r = 1, \ldots, R$) where variable $r$ has $C_r$ categories, coded as $c = 1, \ldots, C_r$. Underlying each ordered categorical response variable $y_r$ we assume a continuous latent variable $z_r$. We model these latent variables as
\[
z_{ir} = \theta_{ir} + \epsilon_{ir},
\]
where $\theta_{ir}$, the structural part of the model, is geometrically defined in $S$ dimensions. When using PCA we define
\[
\theta_{ir}  = \langle \bm{u}_i, \bm{v}_r \rangle = \sum_{s=1}^S u_{is}v_{rs},
\]
with $\bm{u}_i$ the principal scores and $\bm{v}_r$ the loadings, whereas 
\[
\theta_{ir}  = -d(\bm{u}_i, \bm{v}_r) = -\sqrt{\sum_{s=1}^S (u_{is}^2 + v_{rs}^2 - 2u_{is}v_{rs})},
\]
with $\bm{u}_i$ the ideal points for participant $i$ and $\bm{v}_r$ the location for variable $r$ when MDU is used. We denote the two models by cumulative logistic PCA (CLPCA) and cumulative logistic MDU (CLMDU).  

When predictor variables are available for the participants, the coordinates of the principal scores or ideal points can be restricted to be a linear, additive function of these predictor variables, that is $\bm{u}_i = \bm{B}'\bm{x}_i$ with $\bm{B}$ a $P \times S$ matrix. Within the PCA context, we obtain the reduced rank regression model (CLRRR). For MDU, we obtain a restricted multidimensional unfolding (CLRMDU). 

We assume the $\epsilon_{ir}$ to be independent and identically distributed error terms following a cumulative logistic distribution. The probability density function of the logistic distribution equals 
\[
f(\eta) = \frac{\exp(-\eta)}{[1 + \exp(-\eta)]^2} \ \mathrm{for} \ \eta \in \ (-\infty, \infty),
\]
such that its logarithm is
\[
\log f(\eta) = -\eta - 2\log[1 + \exp(-\eta)].
\]
The cumulative density function of this distribution equals
\[
F(\eta) = \frac{1}{1 + \exp(-\eta)} \ \mathrm{for} \ \eta \in \ (-\infty, \infty).
\]

It follows that 
\[
F^{-1}\left(P(y_{ir} \leq c) \right) = \log \left( \frac{P(y_{ir} \leq c)}{P(y_{ir} > c)} \right) = m_{r_c} -  \theta_{ir},
\]
where, similar to the proportional odds regression model, the thresholds ($m_{r_c}$) are \emph{category specific} but the structural part ($\theta_{ir}$) of the model is \emph{variable specific}. 

In this section, we introduced a framework of four models for multidimensional data analysis of ordinal response variables. We first distinguished between models for dominance and proximity variables. Within each of these two types, we distinguish between models with or without predictor variables. 

\subsection{Properties of Cumulative Logistic Models}\label{sec:properties}

Let us consider two subjects with locations $\bm{u}_1$ and $\bm{u}_2$. The cumulative log-odds ratio for response variable $r$ is defined as
\[
\tau = \log \left( \frac{P(y_{1r} \leq c)}{P(y_{1r} > c)} \right) -  \log \left( \frac{P(y_{2r} \leq c)}{P(y_{2r} > c)} \right). 
\]
With a PCA (or RRR) parameterisation, $\tau$ can be written as
\[
\tau = m_{r_c} - \langle \bm{u}_1, \bm{v}_r \rangle - (m_{r_c} - \langle \bm{u}_2, \bm{v}_r \rangle) = \left\langle (\bm{u}_2 - \bm{u}_1), \bm{v}_r \right\rangle, 
\]
which does not depend on $c$. This shows that CLPCA (and CLRRR) make a \emph{proportional odds assumption}, for a given change in the positions for the subjects all the cumulative log-odds for variable $r$ change with the same amount. Furthermore, if we define $\bm{u}_2$ as $\bm{u}_2 = \bm{u}_1 + \delta(\bm{u})$, such that $\delta(\bm{u})$ is a shift from one position to another, then we may write
\[
\tau = \left\langle \delta(\bm{u}), \bm{v}_r \right\rangle, 
\]
which shows that it does not matter where in the Euclidean space this shift happens, the cumulative log-odds ratio remains constant for constant $\delta( \bm{u}) $. Consider participants 1 and 2, with coordinates $\bm{u}_1 = (0,0)$ and $\bm{u}_2 = (1,0.5)$. The estimated cumulative log-odds ratio for these two participants is the same as for participants 3 and 4 with coordinates $\bm{u}_3 = (-5,3)$ and $\bm{u}_4 = (-4, 3.5)$, as the difference between these pairs of coordinates is the same. 

When predictor variables are used in the analysis, we constrain the coordinates to be linear combinations of those variables, that is $\bm{u}_i = \bm{B}'\bm{x}_i$. Of interest is then the comparison between two persons that differ one unit in one of the predictor variables and have equal values for the other predictors. Say, the $p$-th predictor variable increases by a unit, such that, $\delta(\bm{u}) = \bm{B}'(\bm{x}_2 - \bm{x}_1) = \bm{b}_p$, then the cumulative log-odds increase by $\left\langle \bm{b}_p, \bm{v}_r \right\rangle$. 

Below in Section \ref{sec:biplots}, we will describe biplots for the interpretation of our multidimensional models. Cumulative logistic reduced rank regression models can also be interpreted numerically, similar to the regression weights in a proportional odds model. With reduced rank coefficient matrix $\bm{A} = \bm{BV}'$, each column of this matrix represents a change in cumulative log odds for the corresponding response for unit increase in the predictor. 


For cumulative logistic MDU and restricted MDU, there is a nonlinear, distance relationship. The cumulative log-odds ratio becomes
\[
\tau = d(\bm{u}_1,\bm{v}_r) - d(\bm{u}_2,\bm{v}_r), 
\]
again not depending on $c$, only on the distances. This is again the proportional odds assumption, but in contrast with the PCA parameterization, however, the changes in cumulative log odds are not constant for changes in $\bm{u}$. That is, using the example with participants 1, 2, 3, and 4, described above again, the cumulative log-odds ratio for participants 1 and 2 is not equal to that of participants 3 and 4. Similarly, unit changes in one of the predictors do not lead to a constant change in cumulative log-odds (i.e., the estimated $\tau$ is different for two participants with predictor values 0 and 1 compared to two participants with predictor values 2 and 3, say). Furthermore, unit changes in one of the predictors (say, $x_1$) lead to different changes in $\tau$ for participants with varying values on the other predictors. Although the relationship of the predictors is additive in defining the positions of the participants in the biplot, the relationship between predictors is not additive when looking at the effect on the response variables. Results can therefore not be represented numerically and we have to rely on the biplot visualizations described in detail below. 

\subsection{Model Selection}

Assuming conditional independence between the response variables given the representation in low dimensional space, we will estimate the models by maximizing the likelihood (see Section \ref{sec:estimation}, where we derive an algorithm). For selecting a model, that is, finding the optimal dimensionality and selecting a set of predictor variables, likelihood based statistics can be used. We propose to first select an optimal dimensionality, including all predictor variables if available, and thereafter select an optimal set of predictors. For dimensionality selection information criteria (AIC, BIC) can be used. After fixing the dimensionality, we can either use information criteria or likelihood ratio statistics for inference about the predictor variables. In the remainder of the paper, we will use information criteria. 

For the AIC and BIC we need the number of parameters. In all our models, we have the threshold parameters, the number of which is $\sum_r (C_r - 1)$.  
For PCA and RRR, the number of parameters in the structural part is $(N + R - S)S$ and $(P + R - S)S$, respectively. \cite{mukherjee2015degrees} showed, in the context of linear reduced rank models, that these numbers are naive estimates. For linear reduced rank model better estimates are available, but this theory is not extended (yet) to other types of models, such as ordinal response variables. For MDU and RMDU, the number of parameters in the structural part is $(N+R)S - S(S-1)/2$ and $(P+R)S - S(S-1)/2$, respectively. 

\subsection{Biplots}\label{sec:biplots}

In this section, we will discuss biplots for the visualization of the models results. These biplots are most valuable for two-dimensional solutions, but can also be used for visualizaztion of pairs of dimensions in case of higher dimensional solutions. We first discuss biplots for CLPCA and CLMDU. Afterwards, we discuss the case when predictor variables are available for the analysis. In that case, the biplots are extended with extra information about the predictor variables.

\subsubsection{CLPCA biplots}

\begin{figure}
    \centering
    \begin{minipage}{.5\textwidth}
        \centering
        (a)\newline
        \includegraphics[width=0.9\textwidth]{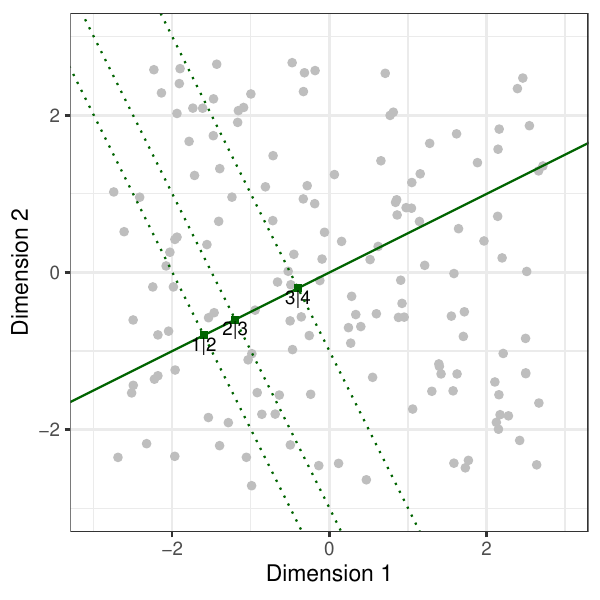}
    \end{minipage}%
    \begin{minipage}{0.5\textwidth}
        \centering
        (b)\newline
        \includegraphics[width=0.9\textwidth]{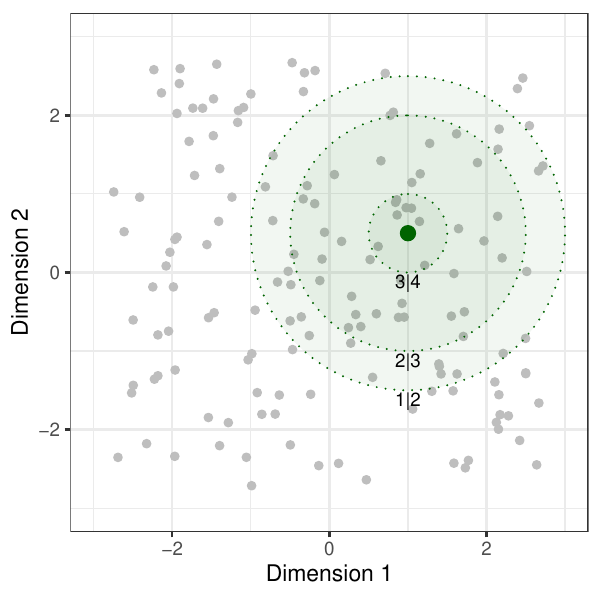}
    \end{minipage}
    \caption{Biplot representations for CLPCA (left) and CLMDU (right) for a single response variable. Variable markers for cumulative probabilities are added. Grey points represent observations. On the left, the green solid line represents the variable axis with markers indicating the estimated thresholds. The dotted lines indicate decision regions for the categories of the response variable. On the right, the green point represents the response variable. The circles represent decision boundaries, where outside the circle the first category of the label is preferred and inside the circle the second category of the label.}
    \label{fig:bgeom}
\end{figure}

Biplots \citep{gabriel1971biplot, gower1996biplots, gower2011understanding} are useful displays for the results of a PCA, especially for two-dimensional solutions. We will now discuss the geometry of the two-dimensional biplot for Cumulative Logistic PCA. Like a usual PCA biplot, observations are shown as points, and variables are shown by axes. The coordinates of the points are given by the estimated $\bm{u}_i$. The variable axes are straight lines though the origin with direction $\frac{v_{r2}}{v_{r1}}$. In Figure \ref{fig:bgeom}a, we present a simplified biplot where the observations are shown by grey dots and there is a single variable axis (solid line). For this variable, $\bm{v}_r = [1, 0.5]'$ and suppose $m_{r_c} = (-2.0, -1.5, -0.5)$ for $c = 1, 2$ and $3$, respectively, i.e., thresholds for a four-point response scale. Estimated values for the response of an individual can be obtained by projecting the point representing this individual onto the variable axis. Subjects positioned in the lower left corner have lower expected values for the response, while subjects in the upper right corner have higher expected values, projecting higher onto the variable axis. To further increase interpretation and provide numerical values for the expected value, \cite{gower1996biplots} suggest to add labeled markers to the variable axis. For CLPCA there are several possibilities.

\cite{vicente2014logistic} suggest to add markers based on the largest estimated a posteriori probabilities. That is, for every point on the variable axis the probability for each response class is computed. At specific locations on the variable axis there are points where two categories, say $c$ and $c'$, jointly have the highest probability. These points are marked as with $c|c'$. They \nocite{vicente2014logistic} note that in some cases the probability of one or several categories are never higher than the probability of the other categories. When, for example, category 2 is such a `hidden' category, the marker will be ``1|3''. 

In the context of the proportional odds model, \cite{anderson1981regression} suggest to make predictions based on the underlying latent variable. Following this suggestion, we propose to add markers based on the underlying latent variable and the estimated thresholds. The predicted value of the latent variable is
\[
\hat{z}_{ir} = \langle \hat{\bm{u}}_i, \hat{\bm{v}}_r \rangle. 
\]
This inner product is constant ($\mu$, say) for all points on a line projecting at the same location of the variable axis, that is, a line orthogonal to the variable axis. Therefore, the point of projection may be calibrated by labelling this point with the value $\mu$. This value also applies to the point of projection itself, which is $\lambda\bm{v}_r$. For the point $\lambda\bm{v}_r$ to be calibrated $\mu$, it must satisfy
\[
\lambda\bm{v}_r'\bm{v}_r = \mu,
\]
so that $\lambda = \mu / \bm{v}_r'\bm{v}_r$ and the coordinates of the point on the variable axis that is calibrated with a value of $\mu$ are $\mu\bm{v}_r /(\bm{v}_r'\bm{v}_r) $. As such, we would have the markers expressing values of the underlying latent continuous variable, but the interest lies in the observed ordinal response variable. The estimated response is $\hat{y} = c$ if $m_{r_{c-1}} \leq \hat{z} < m_{r_{c}}$. Therefore, markers can be based on the estimated thresholds. These markers indicate the transition points between adjacent categories. The coordinates of the marker point are given by $\hat{m}_{r_c} \bm{v}_r /(\bm{v}_r'\bm{v}_r) $ and these can be labeled by 1|2, 2|3, and so forth. The application of these markers is illustrated in Figure \ref{fig:bgeom}a. An advantage of these markers over the ones based on posterior probabilities is that each threshold is represented.  

Based on these markers, the two-dimensional space can be partitioned into $C_r$ areas by drawing $C_r - 1$ decision lines orthogonal on the variable axis and through the marker points, these are represented by the dotted lines in Figure \ref{fig:bgeom}a. In an usual biplot, several response variables are represented by variable axes jointly partitioning the two-dimensional space in open and closed regions that each represent a particular response profile.  

\subsubsection{CLMDU biplots}

In MDU biplots, both the observations and the variables are shown by points in the two-dimensional space. The closer an observation to the variable point the higher the probability of a high response. To represent the ordinal nature of the response variable into the biplot, remember that we have
\[
\log \left( \frac{P(y_{ir} \leq c)}{P(y_{ir} > c)} \right) = m_{r_c} +  d(\bm{u}_i, \bm{v}_r),
\]
so that 
\[
\log \left( \frac{P(y_{ir} > c)}{P(y_{ir} \leq c)} \right) = -m_{r_c} - d(\bm{u}_i, \bm{v}_r) =  a_{r_c} - d(\bm{u}_i, \bm{v}_r). 
\]
It follows that we can add circles to the biplot with center $\bm{v}_r$ and radius $a_{r_c}$, such that for points inside this circle the probability for responding higher than $c$ is larger than 0.5 while outside the circle this probability is smaller or equal to 0.5. Every variable point is therefore accompanied with $C_r - 1$ circles representing the different probabilities. We illustrate the threshold circles in Figure \ref{fig:bgeom}b, where again $\bm{v}_r = [1, 0.5]'$ and $m_{r_c} = (-2.0, -1.5, -0.5)$. From the $m_{r_c}$ we obtain $a_{r_c} = (2.0, 1.5, 0.5)$. We see again that the two dimensional space is partitioned in several areas. These areas are now defined by the circles. Withing the smallest circle around the response variable, the participants are predicted to scores highest. Around the inner circle, we have regions in the form of tyres. Within such a band, participants are predicted to have the same score on the response variable. In some cases the radius can be negative (when $m_{r_c}$ is positive), indicating that nowhere in the two-dimensional space a cumulative probability is larger than a half, and the circle is not drawn. 

In an usual biplot, several response variables are represented.
Each of the points representing a response variable is surrounded by a set of circles. Circles of different response variables cross, creating regions that each represent a particular response profile. 

\subsection{Restricted Models}

In case we have predictor variables available for the observations, the principal scores or ideal points ($\bm{u}_i$) are defined to be linear combinations of the predictor variables, that is $\bm{u}_i = \bm{B}'\bm{x}_i$. To include the predictor variables in the biplot, we distinguish between numerical and categorical predictor variables.  Numerical predictor variables can be included in the two-dimensional biplot as straight lines through the origin and with direction $b_{p2}/b_{p1}$. The positions of the observations can be obtained from the predictor variable axes by the process of \emph{interpolation}, as outlined by \cite{gower1996biplots}. The interpolation process is similar to vector addition, that is, we create vectors starting in the origin and along the variable axes to the observed predictor value for each predictor variables. To obtain the position of the participant, we have to add these vectors. 

Categorical predictor variables, are recoded into dummy variables, where one of the categories is chosen as a reference category. In the biplot representation, we use points instead of variable axes for such predictor variables. The position of the reference category is in the origin of the low dimensional space, whereas the other categories are positioned at their corresponding estimates in $\bm{B}$. We can consider categorical predictor variables as ``jumps''. When the categorical predictor is equal to the reference category, no jump is made. When the categorical predictor variable is equal to another category, a jump from one position to another is made. This jump changes the location of variable axes for the numerical predictor variables. 

Variable axes for the numerical predictor variables and points for the categorical predictor variables are added to the biplots for CLPCA or CLMDU. For interpretation, the relationship between a predictor variable and a response variable is of interest. In the CLRRR biplots, for numerical predictor variables such a relationship is given by the angle between a predictor variable axis and that of a response variable (see Section \ref{sec:properties}). A sharp angle indicates a strong relationship, while an obtuse angle indicates the absence of a relationship. Furthermore, for every point along the predictor variable axis, a projection onto the response variable axis can be made to obtain a predicted value. For categorical predictor, the point representing a category can be projected  onto the response variable axis can be made to obtain a predicted value.

For CLRMDU biplots, the interpretation of predictor-response relationships is more involved, because these are single-peaked where with increasing values of a predictor first the response goes up and afterwards down again. Furthermore, although the effect of predictor variables is additive for obtaining the position of an observation (i.e., the ideal point $\bm{u}_i$), this additivity does not translate to the relationship towards the response variable. 

\section{Maximum Likelihood Estimation}\label{sec:estimation}

Assuming a multinomial distribution of the response variables, the \emph{observed data negative log-likelihood} is 
\[
\mathcal{L}_o(\bm{\theta}, \bm{m}) = - \sum_i\sum_r \mathds{1}(y_{ir} = c) \log \pi_{irc},
\]
where $\pi_{irc} = F(m_{r_c} - \theta_{ir}) - F(m_{r_{c-1}} - \theta_{ir})$, and $\mathds{1}()$ is an indicator function of its argument, $\bm{\theta}$ collects all the structural parameters and $\bm{m}$ collects all the threshold parameters. 

In this section, we will develop an Expectation Majorization Minimization (EMM) algorithm to minimize the negative loglikelihood. This algorithm is a combination of the EM-algorithm often used for latent variable models \citep{mclachlan2007algorithm} and the MM-algorithm \citep{hunter2004tutorial, heiser1995convergent}.

We start by formulating the complete data negative loglikelihood, take the conditional expectation of this function in the E-step, and find a majorization function that can easily be minimized. The majorization function turns out to be a least squares function, so that in the inner loop of the algorithm, well known updating steps from least squares theory can be used. 

\subsection{Estimation of Structural part}

The \emph{complete data negative log-likelihood} (CDNLL) is defined for the latent responses, that is 
\[
\mathcal{L}_c(\bm{\theta}) = - \sum_i\sum_r \log f(z_{ir} - \theta_{ir}), 
\]
where \(f(\cdot)\) is the probability density function of the logistic distribution (see Section 2). 
A second-order Taylor expansion of the complete data negative log-likelihood around current values $\tilde{\bm{\theta}}$ is 
\[
\mathcal{L}_c(\bm{\theta}) = \mathcal{L}_c(\tilde{\bm{\theta}}) + (\bm{\theta} - \tilde{\bm{\theta}})' \frac{\partial \mathcal{L}_c(\tilde{\bm{\theta}})}{\partial \bm{\theta}}
+ \frac{1}{2} (\bm{\theta} - \tilde{\bm{\theta}})' \frac{\partial^2 \mathcal{L}_c(\tilde{\bm{\theta}})}{\partial \bm{\theta} \partial \bm{\theta}'} (\bm{\theta} - \tilde{\bm{\theta}}).
\]

In the E-step, the expectation of the complete data negative log-likelihood is obtained, that is 
\[
\mathbb{E}(\mathcal{L}_c(\bm{\theta})) = \mathbb{E}(\mathcal{L}_c(\tilde{\bm{\theta}})) + (\bm{\theta} - \tilde{\bm{\theta}})' \mathbb{E}\left(\frac{\partial \mathcal{L}_c(\tilde{\bm{\theta}})}{\partial \bm{\theta}}\right)
+ \frac{1}{2} (\bm{\theta} - \tilde{\bm{\theta}})' \mathbb{E}\left(\frac{\partial^2 \mathcal{L}_c(\tilde{\bm{\theta}})}{\partial \bm{\theta} \partial \bm{\theta}'}\right) (\bm{\theta} - \tilde{\bm{\theta}})
\]
As the expectation of a sum is the sum of expectations, we can write
\[
\mathbb{E}(\mathcal{L}_c(\bm{\theta})) = \sum_i \sum_r \mathbb{E}(\mathcal{L}_c(\theta_{ir})),
\]
with
\[
\mathbb{E}(\mathcal{L}_c(\theta_{ir})) = 
\mathbb{E}(\mathcal{L}_c(\tilde{\theta}_{ir})) + (\theta_{ir} - \tilde{\theta}_{ir}) \mathbb{E}\left(\frac{\partial \mathcal{L}_c(\tilde{\theta}_{ir})}{\partial \theta_{ir}}\right)
+ \frac{1}{2} (\theta_{ir} - \tilde{\theta}_{ir}) \mathbb{E}\left(\frac{\partial^2 \mathcal{L}_c(\tilde{\theta}_{ir})}{\partial^2 \theta_{ir}}\right) (\theta_{ir} - \tilde{\theta}_{ir}).
\]

An upper bound for the (expectation of the) second derivative is given by $1/4$ such that 
\[
\mathbb{E}(\mathcal{L}_c(\theta_{ir})) \leq 
\mathbb{E}(\mathcal{L}_c(\tilde{\theta}_{ir})) + (\theta_{ir} - \tilde{\theta}_{ir}) \mathbb{E}\left(\frac{\partial \mathcal{L}_c(\tilde{\theta}_{ir})}{\partial \theta_{ir}}\right)
+ \frac{1}{8} (\theta_{ir} - \tilde{\theta}_{ir}) (\theta_{ir} - \tilde{\theta}_{ir}) = \mathcal{M}(\theta_{ir}|\tilde{\theta}_{ir}),
\]
where $\mathcal{M}(\theta_{ir}|\tilde{\theta}_{ir})$ is the majorization function. Let us define \( p_{ir} = \frac{1}{1 + \exp(-z_{ir} + \theta_{ir})}\) such that  \( \log f(z_{ir} - \theta_{ir}) = \log p_{ir}(1 - p_{ir})\). The partial derivative is
\[
\frac{\partial \mathcal{L}_c(\tilde{\theta}_{ir})}{\partial \theta_{ir}} = - \frac{\partial \log f(z_{ir} - \theta_{ir})}{\partial \theta_{ir}}  = 1 - 2p_{ir}.
\] 
We need the expectation of $p_{ir}$ to find the expectation of this partial derivative. Following \cite{jiao2016high}, we derived the following closed form expressions (we use $p$, $y$ and $\theta$ instead of $p_{ir}$, $y_{ir}$ and $\theta_{ir}$ for readibility):
\begin{eqnarray*}
\mathbb{E}(p| y, \theta, m)  = \left\{ \begin{array}{ll} 
\left[\frac{\exp(2m_{y} - 2\theta)} {2[\exp(m_{y} - \theta) + 1]^2}\right] / F(m_{y} - \theta)		& \mathrm{if}\ y = 1\\[4pt]
\left[\frac{2\exp(m_{(y-1)} - \theta) + 1} {2[\exp(m_{(y-1)} - \theta) + 1]^2} - \frac{2\exp(m_{y} - \theta)} {2[\exp(m_{y} - \theta) + 1]^2}\right]	/ \left(F(m_{y} - \theta)	- F(m_{(y-1)} - \theta)\right)	& \mathrm{if}\ 2\leq y < C \\[4pt]
\left[\frac{2\exp(m_{(y-1)} - \theta) + 1} {2[\exp(m_{(y-1)} - \theta) + 1]^2}\right] / \left( 1 - F(m_{(y-1)} - \theta) \right)			& \mathrm{if}\ y = C
\end{array}\right . 
\end{eqnarray*} 
The expectation has to be evaluated at the current estimates of $\bm{\theta}$ and $\bm{m}$. Let us denote by $\xi_{ir}$ the expected value of the first derivative, that is
\[
\xi_{ir} = 1 - 2 \mathbb{E}(p_{ir}| y_{ir}, \theta_{ir}, m_r),
\]
so that the majorization function to be minimized is 
\[
\mathcal{M}(\bm{\theta}|\tilde{\bm{\theta}}) = \sum_i \sum_r \mathcal{M}(\theta_{ir}|\tilde{\theta}_{ir}) =  \sum_i \sum_r \mathbb{E}(\mathcal{L}_c(\tilde{\theta}_{ir})) + (\theta_{ir} - \tilde{\theta}_{ir}) \xi_{ir} + \frac{1}{8} (\theta_{ir} - \tilde{\theta}_{ir}) (\theta_{ir} - \tilde{\theta}_{ir}).
\]
Let us simplify this majorization function. Focusing on the individual elements, the first term is a constant ($c_1 = \mathbb{E}(\mathcal{L}_c(\tilde{\theta}_{ir}))$) and therefore we focus on the other terms
\[
\begin{aligned}
\mathcal{M}(\theta_{ir}|\tilde{\theta}_{ir}) &= c_1 +  \xi_{ir}(\theta_{ir} - \tilde{\theta}_{ir}) + \frac{1}{8}(\theta_{ir} - \tilde{\theta}_{ir})(\theta_{ir} - \tilde{\theta}_{ir}) \\
&= c_1 + \xi_{ir}\theta_{ir} - \xi_{ir}\tilde{\theta}_{ir} + \frac{1}{8}(\theta_{ir}^2 + \tilde{\theta}_{ir}^2 -2\theta_{ir}\tilde{\theta}_{ir}) \\
&= c_1 + \frac{1}{8}\theta_{ir}^2 + \xi_{ir}\theta_{ir} -2\frac{1}{8}\theta_{ir}\tilde{\theta}_{ir} - \xi_{ir}\tilde{\theta}_{ir} + \frac{1}{8}\tilde{\theta}_{ir}^2 .
\end{aligned}
\] 
Now, let us define, what we will call \emph{working responses}
\begin{eqnarray}
\lambda_{ir} = \tilde{\theta}_{ir} - 4\xi_{ir}
\label{eq:workingresponses}
\end{eqnarray}
to obtain 
\[
\begin{aligned}
\mathcal{M}(\theta_{ir}|\tilde{\theta}_{ir}) &= c_1 + \frac{1}{8}\theta_{ir}^2  -2\frac{1}{8}\theta_{ir}\lambda_{ir}  + \frac{1}{8}\lambda_{ir}^2 - \frac{1}{8}\lambda_{ir}^2 - \xi_{ir}\tilde{\theta}_{ir} + \frac{1}{8}\tilde{\theta}_{ir}^2 \\
&= c_1+ \frac{1}{8}(\theta_{ir} - \lambda_{ir})^2 - \frac{1}{8}\lambda_{ir}^2 - \xi_{ir}\tilde{\theta}_{ir} + \frac{1}{8}\tilde{\theta}_{ir}^2.
\end{aligned}
\] 
Define $ c_2 = c_1 - \frac{1}{8}\lambda_{ir}^2 - \xi_{ir}\tilde{\theta}_{ir} + \frac{1}{8}\tilde{\theta}_{ir}^2$, a constant with respect to $\theta_{ir}$, so that we can write
\[
\mathbb{E}(\mathcal{L}_{ir}(\theta_{ir})) \leq \mathcal{M}(\theta_{ir}|\tilde{\theta}_{ir}) = \frac{1}{8} (\theta_{ir} - \lambda_{ir})^2 + c_2.  
\] 
Now collecting all terms into a single function, we obtain
\[
\mathbb{E}(\mathcal{L}_c(\bm{\theta})) \leq \mathcal{M}(\bm{\theta}|\tilde{\bm{\theta}}) = \frac{1}{8} \sum_i \sum_r (\theta_{ir} - \lambda_{ir})^2 + c_2, 
\] 
a least squares function with the working responses $\lambda_{ir}$. In the following four subsections, we work out this least squares loss function for the four different definitions of the structural part ($\theta_{ir}$). 

\subsubsection{PCA parametrisation of the Structural part}

Remember that $\theta_{ir} =  \langle \bm{u}_{i}, \bm{v}_{r} \rangle$, so that the loss function equals
\[
\sum_i \sum_r (\lambda_{ir} - \bm{u}_{i}' \bm{v}_{r})^2,
\]
which can be written in matrix algebra terms as
\[
\| \bm{\Lambda} - \bm{U}\bm{V}'\|^2. 
\]
We have to find a reduced rank approximation of the matrix $\bm{\Lambda}$ with elements $\lambda_{ir}$. \cite{eckart1936approximation} showed that this can be done using a singular value decomposition 
\[
\bm{\Lambda} = \bm{P}\bm{\Phi}\bm{Q}'
\]
and defining the updates 
\begin{eqnarray}
\bm{U}^+ &=& \sqrt{N} \bm{P}_S \label{eq:PCAupdateU} \\
\bm{V}^+ &=& \frac{1}{\sqrt{N}} \bm{Q}_S\bm{\Phi}_S, \label{eq:PCAupdateV}
\end{eqnarray} 
where $\bm{P}_S$ and $\bm{Q}_S$ are the first $S$ singular vectors, and $\bm{\Phi}_S$ is the $S \times S$ diagonal matrix with the largest singular values. 

\subsubsection{RRR parametrisation of the Structural part}

Compared to the PCA parametrisation, we impose the constraint that $\bm{u}_i = \bm{B}'\bm{x}_i$. Therefore, in each iteration, we have to minimize
\[
\| \bm{\Lambda} - \bm{XB}\bm{V}' \|^2
\]
over the parameters $\bm{B}$ and $\bm{V}$. Updates of $\bm{B}$ and $\bm{V}$ can be obtained from a generalized singular value decomposition of the matrix
\[
(\bm{X}'\bm{X})^{-1}\bm{X}'\bm{\Lambda}
\]
in the metrics $(\bm{X}'\bm{X})$ and $\bm{I}$ \citep{takane2013constrained}. Let  
\[
(\bm{X}'\bm{X})^{-\frac{1}{2}}\bm{X}'\bm{\Lambda} = \bm{P}\bm{\Phi}\bm{Q}'
\]
be the usual SVD. The updates are defined as
\begin{eqnarray}
\bm{B}^+ &=& \sqrt{N} (\bm{X}'\bm{X})^{-\frac{1}{2}} \bm{P}_S \label{eq:RRRupdateB}\\ 
\bm{V}^+ &=& \frac{1}{\sqrt{N}} \bm{Q}_S\bm{\Phi}_S, \label{eq:RRRupdateV}
\end{eqnarray}
where $\bm{P}_S$ and $\bm{Q}_S$ are the first $S$ singular vectors, and $\bm{\Phi}_S$ is the $S \times S$ diagonal matrix with the largest singular values.

\subsubsection{MDU parametrisation of the Structural part}

The loss function in every iteration is
\[
 \frac{1}{8} \sum_i \sum_r (\theta_{ir} - \lambda_{ir})^2 + c_2. 
\] 
where $\theta_{ir} = -d(\bm{u}_i, \bm{v}_r)$ with $\bm{u}_i$ and $\bm{v}_r$ the parameters. This loss function can be rewritten as 
\[
\sum_{i} \sum_r w_{ir}(\delta_{ir} - d(\bm{u}_i, \bm{v}_r))^2 ,
\] 
where \(\delta_{ir} = -\lambda_{ir}\) and $w_{ir} = 1$, which is the usual raw STRESS function often used
in multidimensional scaling and unfolding. \cite{deleeuw1977applications} and \cite{deleeuw1977convergence} proposed the SMACOF algorithm for minimization of this STRESS function for multidimensional scaling. The SMACOF algorithm is itself an MM algorithm. Convergence properties of this algorithm are described by \cite{deleeuw1988convergence}. 
\cite{heiser1981unfolding, heiser1987joint} showed that multidimensional unfolding can be considered a special case of multidimensional scaling. Subsequently, he developed the SMACOF algorithm to deal with rectangular proximity matrices. Advances in the algorithm are described in \cite{busing2010advances}. An elementary treatment of the algorithm for multidimensional scaling can be found in Chapter 8 of \cite{borg2005modern} and for multidimensional unfolding in Chapter 14. The critical difference with the usual loss function is that the dissimilarities \(\delta_{ir}\) might be negative. \cite{heiser1991generalized} showed a way to deal with negative dissimilarities in multidimensional scaling. The line of thought of Heisers contribution is that two majorizing functions are defined: one for the case that the dissimilarity is
positive and one for the case that the dissimilarity is negative. It turns out that the new algorithm is a simple adaptation of the standard
SMACOF algorithm, where only some elements of two matrices ($\bm{A}$ and $\bm{W}$, see below) are defined differently, depending on the sign of the dissimilarity. \cite{derooijbusing2022} and \cite{derooijwoestenburgbusing2022}  adapted Heiser's algorithm for the multidimensional unfolding case. Here, we will follow that approach. We will only show the updating equations, for the derivation of these equations we refer to the above papers. 

Define matrix \(\bm{A}\) with elements \(\{a_{ir}\}\) as follows 
\begin{eqnarray*}
a_{ir} = \left\{ \begin{array}{ll} 
w_{ir}\delta_{ir} / d(\bm{u}_i, \bm{v}_r) & \mathrm{if}\ \delta_{ir} \geq 0 \ \mathrm{and} \ d(\bm{u}_i, \bm{v}_r) > 0 \\
0                                          & \mathrm{else}
\end{array}\right. .
\end{eqnarray*}

Furthermore, redefine the weight matrix \(\bm{W}\) with
elements \(\{w_{ir}\}\) as 
\begin{eqnarray*}
w_{ir} = \left\{ \begin{array}{ll} 
w_{ir}                                          & \mathrm{if}\ \delta_{ir} \geq 0\\
\left[ w_{ir}(d(\bm{u}_i, \bm{v}_r) + |\delta_{ir}|)\right] / d(\bm{u}_i, \bm{v}_r)      & \mathrm{if}\ \delta_{ir} < 0\  \mathrm{and}\ d(\bm{u}_i, \bm{v}_r) > 0 \\
\left[ w_{ir}(\epsilon + \delta_{ir}^2) \right] /\epsilon  & \mathrm{if}\ \delta_{ir} < 0\  \mathrm{and}\ d(\bm{u}_i, \bm{v}_r) = 0
\end{array}\right. , 
\end{eqnarray*} 
where \(\epsilon\) is a small constant. Note that the
matrices \(\bm{W}\) and \(\bm{A}\) change from iteration
to iteration.

Let us now define \(\bm{R} = \mathrm{diag}(w_{i+})\),
\(\bm{C} = \mathrm{diag}(w_{+r})\),
\(\bm{P} = \mathrm{diag}(a_{i+})\), and
\(\bm{Q} = \mathrm{diag}(a_{+r})\), where the + in the subscript means taking the sum over the replaced index, that is,  $w_{i+} = \sum_{r=1}^R w_{ir}$. With these matrices the update for $\bm{U}$ is
\begin{eqnarray}
\bm{U}^+ = \bm{R}^{-1} \left(\bm{P}\bm{U}- \bm{A}\bm{V} + \bm{W}\bm{V} \right) \label{eq:MDUupdateU}
\end{eqnarray}
and for $\bm{V}$ the update is
\begin{eqnarray}
\bm{V}^+ = \bm{C}^{-1} \left(\bm{Q}\bm{V}- \bm{A}'\bm{U}^+ + \bm{W}'\bm{U}^+\right). \label{eq:MDUupdateV}
\end{eqnarray}
These updates are the same as in the standard least squares unfolding
algorithm \citep[see][pp. 176, 183-187]{busing2010advances}, where only the
definitions of \(\bm{A}\) and \(\bm{W}\) are changed.


\subsubsection{RMDU parametrisation of the Structural part}

When predictor variables are available, we constrain $\bm{U} = \bm{XB}$ and we need to estimate $\bm{B}$ instead of $\bm{U}$. We 
can use the algorithm of MDU described in the previous section and replace the updating equation for $\bm{U}$ (i.e., Equation \ref{eq:MDUupdateU}) with an updating equation for $\bm{B}$, that is  
\begin{eqnarray}
\bm{B}^+ = \left(\bm{X}'\bm{R}\bm{X} \right)^{-1} \left[\bm{X}' \left(\bm{P}\bm{U}- \bm{A}\bm{V}\right) + \bm{X}'\bm{W}\bm{V}\right] \label{eq:RMDUupdateB}
\end{eqnarray}
and before updating $\bm{V}$ using Equation \ref{eq:MDUupdateV}, we compute $\bm{U}^+ = \bm{XB}^+$.


\subsection{Estimation of thresholds}

To update the threshold parameters for each response variable we cannot use the complete data negative log-likelihood. However, we can use the default maximum likelihood estimator as in the proportional odds regression model. For this estimator $\bm{y}_r$ is the response variable and we use $\bm{\theta}_r$ as an offset (i.e., a predictor variable with regression weight fixed to 1) without any further predictor variables. This gives maximum likelihood estimates of the intercepts or thresholds for response variable $r$. We repeat the procedure for each response variable.  

\subsection{Remarks on Algorithms}

The algorithms as outlined above monotonically converge to a local minimum of the negative log likelihood function. For the models based on the inner product (PCA and RRR) this local minimum is also the global minimum. For models based on the distance representation, however, local optima occur. To deal with these local optima, starting values near the global minimum might help, such as given by, for example, correspondence analysis. As such a start does not guarantee to find the global minimum, supplementary multiple random starts are advised. 

\subsection{Algorithm schemes}

To summarize, we give algorithm schemes in four algorithm boxes. Algorithm 1 shows the procedure for CLPCA, algorithm 2 for CLRRR, algorithm 3 for CLMDU, and algorithm 4 for CLRMDU. 

\begin{minipage}{0.46\textwidth}
\begin{algorithm}[H]\label{alg:clpca}
  \scriptsize
  \caption{CLPCA algorithm.}
  \begin{algorithmic}[1]
  \Procedure{clpca}{$\bm{Y}, \bm{m} , \bm{U}, \bm{V}$}
  \State Predefine: maxouter, $\epsilon$
  \State Compute structural part $\bm{\theta}$
  \State Assess $\mathcal{L}_o^0(\bm{\theta}, \bf{m})$
  \For {$t_1 \leftarrow 1,\text{maxouter}$}
    \State Compute working responses $\bm{\Lambda} = \{ \lambda_{ir} \}$ using equation (\ref{eq:workingresponses})
    \State SVD of $\bm{\Lambda}$
    \State Update $\bm{U}$ using equation (\ref{eq:PCAupdateU})
    \State Update $\bm{V}$ using equation (\ref{eq:PCAupdateV})
    \State Update $\bm{m}$ by directly maximizing the likelihood
    \State if $\mathcal{L}_o^{t_1}(\bm{\theta}\bm{m}) - \mathcal{L}_o^{t_1-1}(\bm{\theta}, \bm{m}) < \epsilon$: break
  \EndFor
  \State return($\bm{m}, \bm{U},\bm{V}$)
  \EndProcedure
  \end{algorithmic}
\end{algorithm}
\end{minipage}
\hfill
\begin{minipage}{0.46\textwidth}
\begin{algorithm}[H]\label{alg:clrrr}
  \scriptsize
  \caption{CLRRR algorithm.}
  \begin{algorithmic}[1]
  \Procedure{clrrr}{$\bm{Y}, \bm{X}, \bm{m} , \bm{B}, \bm{V}$}
  \State Predefine: maxouter, $\epsilon$
  \State Compute structural part $\bm{\theta}$
  \State Assess $\mathcal{L}_o^0(\bm{\theta}, \bf{m})$
  \For {$t_1 \leftarrow 1,\text{maxouter}$}
  \State Compute working responses $\bm{\Lambda} = \{ \lambda_{ir} \}$ using equation (\ref{eq:workingresponses})
  \State GSVD of $\bm{\Lambda}$
  \State Update $\bm{B}$ using equation (\ref{eq:RRRupdateB})
  \State Update $\bm{V}$ using equation (\ref{eq:RRRupdateV})
  \State Update $\bm{m}$ by directly maximizing the likelihood
  \State if $\mathcal{L}_o^{t_1}(\bm{\theta}\bm{m}) - \mathcal{L}_o^{t_1-1}(\bm{\theta}, \bm{m}) < \epsilon$: break
  \EndFor
  \State return($\bm{m}, \bm{B},\bm{V}$)
  \EndProcedure
  \end{algorithmic}
\end{algorithm}
\end{minipage}

\begin{minipage}{0.46\textwidth}
\begin{algorithm}[H]\label{alg:clmru}
  \scriptsize
  \caption{CLMDU algorithm.}
  \begin{algorithmic}[1]
  \Procedure{clmru}{$\bm{Y}, \bm{m} , \bm{U}, \bm{V}$}
  \State Predefine: maxouter, maxinner, $\epsilon_1$, $\epsilon_2$
  \State Compute structural part $\bm{\theta}$
  \State Assess $\mathcal{L}_o^0(\bm{\theta}, \bf{m})$
  \For {$t_1 \leftarrow 1,\text{maxouter}$}
   \State Compute working responses $\bm{\Lambda} = \{ \lambda_{ir} \}$ using equation (\ref{eq:workingresponses})
    \State Assess $\mathcal{M}^{0}(\bm{U}, \bf{V})$
    \For {$t_2 \leftarrow 1,\text{maxinner}$}
        \State Update $\bm{U}$ using equation (\ref{eq:MDUupdateU})
        \State Update $\bm{V}$ using equation (\ref{eq:MDUupdateV})
        \State Assess $\mathcal{M}^{t2}(\bm{U}, \bf{V})$
        \State if $\mathcal{M}^{t_2}(\bm{U}, \bm{V}) - \mathcal{M}^{t_2-1}(\bm{U}, \bm{V}) < \epsilon_2$: break
      \EndFor
    \State Update $\bm{m}$ by directly maximizing the likelihood
    \State if $\mathcal{L}_o^{t_1}(\bm{\theta}\bm{m}) - \mathcal{L}_o^{t_1-1}(\bm{\theta}, \bm{m}) < \epsilon_1$: break
  \EndFor
  \State return($\bm{m}, \bm{U},\bm{V}$)
  \EndProcedure
  \end{algorithmic}
\end{algorithm}
\end{minipage}
\hfill
\begin{minipage}{0.46\textwidth}
\begin{algorithm}[H]\label{alg:clrmdu}
  \scriptsize
  \caption{CLRMDU algorithm.}
  \begin{algorithmic}[1]
  \Procedure{clrmdu}{$\bm{Y}, \bm{X}, \bm{m} , \bm{B}, \bm{V}$}
  \State predefine: maxouter, maxinner, $\epsilon_1$, $\epsilon_2$
  \State Compute structural part $\bm{\theta}$
  \State Assess $\mathcal{L}_o^0(\bm{\theta}, \bf{m})$
  \For {$t_1 \leftarrow 1,\text{maxouter}$}
  \State Compute working responses $\bm{\Lambda} = \{ \lambda_{ir} \}$ using equation (\ref{eq:workingresponses})
  \State Assess $\mathcal{M}^{0}(\bm{U}, \bf{V})$
    \For {$t_2 \leftarrow 1,\text{maxinner}$}
    \State Update $\bm{B}$ using equation (\ref{eq:RMDUupdateB})
    \State Update $\bm{V}$ using equation (\ref{eq:MDUupdateV})
    \State Assess $\mathcal{M}^{t2}(\bm{U}, \bf{V})$
    \State if $\mathcal{M}^{t_2}(\bm{U}, \bm{V}) - \mathcal{M}^{t_2-1}(\bm{U}, \bm{V}) < \epsilon_2$: break
    \EndFor
    \State Update $\bm{m}$ by directly maximizing the likelihood
    \State if $\mathcal{L}_o^{t_1}(\bm{\theta}\bm{m}) - \mathcal{L}_o^{t_1-1}(\bm{\theta}, \bm{m}) < \epsilon_1$: break
  \EndFor
  \State return($\bm{m}, \bm{B},\bm{V}$)
  \EndProcedure
  \end{algorithmic}
\end{algorithm}
\end{minipage}

\section{Applications}\label{sec:applications}

In this section, we will show three applications of our modeling framework. In the first application, we analyze part of the data described by \cite{fabbricatore2024} about students responses to exam questions about statistics and relationships with several psychological variables like attitude and anxiety. We will analyze these data using the cumulative logistic reduced rank model. In the second application, we use data from the International Social Survey program to investigate the relationship between environmental attitudes and pro-environmental behaviour. In this analysis, we use both the dominance and proximity models to highlight the differences.  Finally, we show a third application again using data from the International Social Survey program. We use data from seven Likert scales to investigate differences between countries in environmental efficacy. We use the cumulative logistic restricted multidimensional unfolding model for this analysis. 

\subsection{Students' Performance for Statistical Tests}

The subset of data we use for this analysis involves 138 university students and their responses to ten questions in an exam about the application of statistics to certain topics. The ten items are described in detail in Appendix A. Each response is coded as wrong (0), partially correct (1), or correct (2), that is, as a three categories ordinal response. 

Prior to the start of the courses on statistics, several psychological tests were conducted. Besides their answers on the ten items, for each student information is available on their gender (1 = female), age in years, and several measurements of mathematical knowledge and psychological factors, each assessed through validated psychometric instruments. The variables that we use are: 
\begin{itemize}
\item \emph{Mathematical knowledge} measured using the Mathematical Prerequisites for Psychometrics scale (PMP); 
\item \emph{Statistical anxiety} with three scale scores referring to examination anxiety (SASa), interpretation anxiety (SASi), and fear of asking for help (SASf);
\item \emph{Attitudes toward Statistics} with four scales: 
affect (SATSa), cognitive competence (SATSc), value (SATSv), and difficulty (SATSd);
\item \emph{Motivated strategy for learning} with four scales referring to 
self-efficacy (MSLe), test anxiety (MSLt), cognitive strategies (MSLc), and self-regulation (MSLsr);
\item \emph{Academic procrastination} measured using a single scale (APS); 
\item \emph{Academic motivation} also measured using a single scale (AMS);
\item \emph{Student engagement in statistics} measured through three scales:
affective engagement (ENGa), behavioural engagement (ENGb), and cognitive engagement (ENGc).
\end{itemize}
Psychometric properties of these scales are all satisfactory to good, see \cite{fabbricatore2024} for details, where also more detailed descriptions of the scales can be found. The scores on these scales serve, together with gender and age, as predictor variables. The responses to the ten statistical application items serve as response variables. 

As there is no prior information whether the 10 items comprise a unidimensional or multidimensional construct, we will start fitting models in one till three dimensions and select an optimal dimensionality. Subsequently, we will verify which of the predictor variables influence the responses. Fit statistics for the one-, two-, and three dimensional models are shown in Appendix A, where it can be seen that the two-dimensional model has the lowest AIC, while the lowest BIC is obtained for the unidimensional model. We proceed with the two-dimensional model. Leaving out each (set of) predictor variables, we obtain the fit statistics in the lower part of Table A1,  showing that Age, Statistical Anxiety, the Motivated Strategies for Learning Questionnaire, and the Academic Motivation Scale can be left out without significant loss of fit (AIC based conclusion). 

\begin{figure}[ht!]
\begin{center}
\includegraphics[width = 1.1\textwidth]{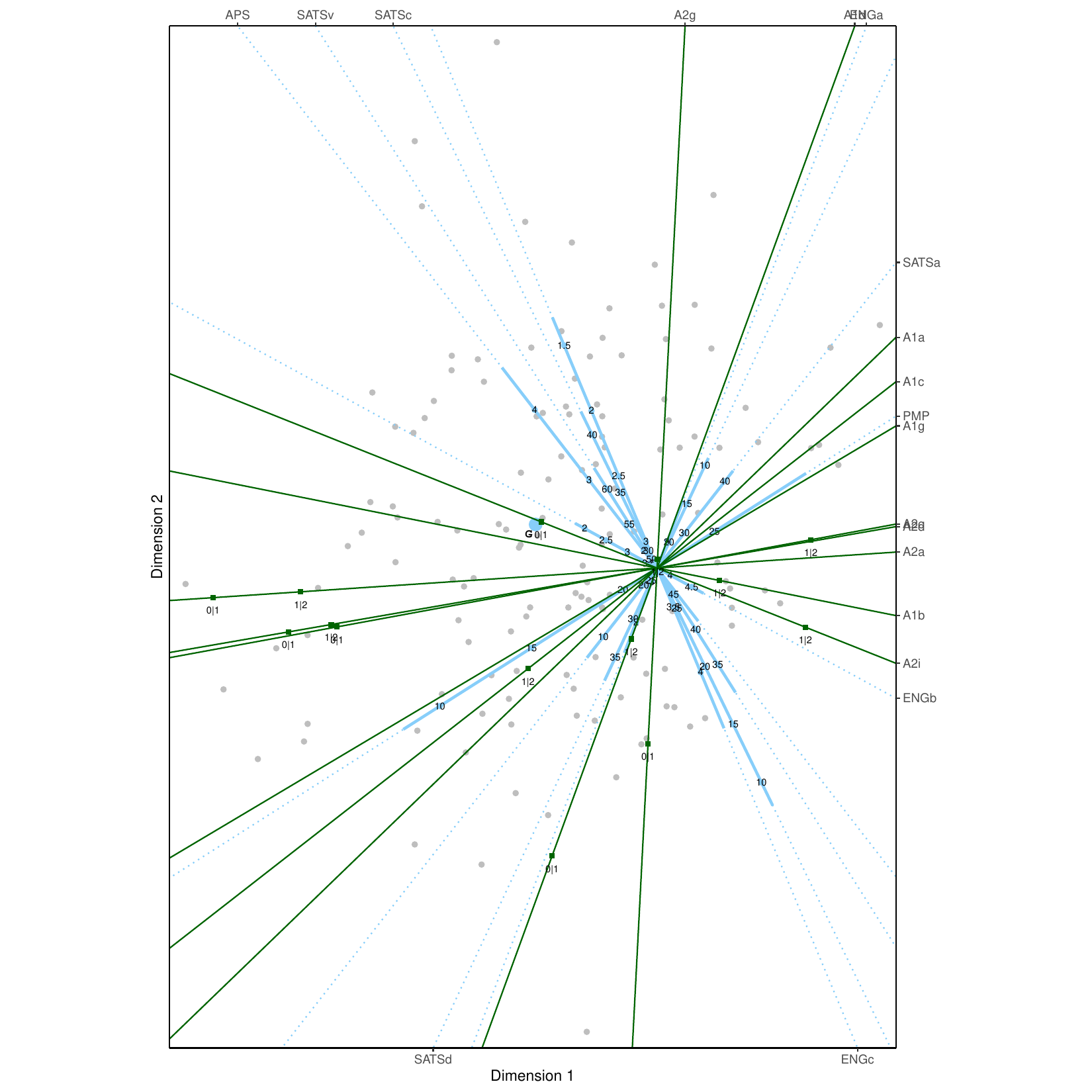}
\caption{Estimated configuration for students data. The dark green lines represent the response variables, the blue lines represent the predictor variables. Variable labels are placed on the positive side of the variables, that are the sides with the largest values. In the upper right corner the labels of ENGa and A1d overlap.}
\label{fig:student}
\end{center}
\end{figure}

The final biplot is shown in Figure \ref{fig:student}. The implied coefficients are shown in Table A2 (Appendix A). The figure and table can be used together to come to a interpretation of the final model. Let us first look at the configuration of the response variables (items A1a, \ldots, A2g) represented by the green solid lines with markers 0|1 and 1|2. By inspecting the direction of the variable axes, we see most responses are strongly correlated, that is, the variable axes have small angles and all point to the right hand side of the figure. A notable exception is item A2g, and to a smaller extend A1d. The general pattern is that students on the right hand side of the figure correctly answer the items, while students on the left hand side often make mistakes. For items A2g and A1d, students in the top of the configuration make the items correct, while at the bottom of the figure they tend to fail. Each variable axis also has two markers (0|1 and 1|2) indicating the difficulty of the item. For response variable A2a, for example, we see that these two markers are strongly to the left. Projecting the students on the axes we see that most fall in the region of a correct answer. This item is relatively easy. For item A2i, however, the marker 1|2 is far to the right and the projection of student points on this axis is rarely in the region of a correct answer. Note that item A2i is also the item with the least correct answers in the data, whereas A2a has many correct responses. 

Now let us further inspect the predictor variables (indicated by the light blue variable axes) and the positions of the students in the configuration. The predictor variables are represented with a variable axis that has a solid part and a dotted part: the solid part indicates the observed range in the data, from the minimum to the maximum observed value, and as such functions as a measure of effect size. The dotted part only extends the variable axis to the edge of the biplot where the variable label is placed. It can be seen that PMP has a long solid variable axis, that is, mathematical knowledge prior to the courses makes large differences. Students who score high on mathematics (PMP) are positioned on the right hand side. It follows that a high PMP score is predictive for answering most items correctly. Similarly, high behavioural engagement (ENGb) and a strong positive feeling about statistics (i.e., a high score on the affect scale SATSa) result in positions more to the right hand side of the biplot, where the model indicates correct responses to the items. 

The predictor variable affective engagement (ENGa) points to the upper right corner, so is a good predictor for a correct response on A1d and A2g. Students who consider statistics to be a difficult topic (SATSd), in contrast, will be on the lower left corner, that is, the variable axis points in the opposite direction. Therefore, \emph{lower scores} on this variable predict a correct response on A1d and A2g. 

Higher scores on academic procrastination (APS) and the value and cognitive competence scales of attitudes towards statistics (SATSv and SATSc, respectively) result in positions more to the top left. These points project low  on most of the response variables, indicating wrong answers,  but high (i.e., correct answers) on A2g and A1d. In contrast,  higher scores on cognitive engagement (ENGc) result in positioning more towards the bottom right, an opposite response pattern compared to academic procrastination (APS) and the two attitudes scales (SATSv and SATSc). Finally, the gender variable (G) points towards the left: When a boy and a girl have equal values for the other predictor variables, girls are positioned more to the left hand side, indicating less favorable answers to the ten statistics items.   

\subsection{ISSP data: Pro-environmental behaviour}\label{sec:peb}

In this application and the next, we use from the  International Social Survey Programme 2020, the Module on Environment \citep{ISSP2020data}. In this first analysis, we focus on the data from Thailand ($N = 1063$) and the responses to four pro-environmental behaviour variables, measured on ordinal scales:
\begin{itemize}
    \item[OUT] \emph{In the last twelve months how often, if at all, have you engaged in any leisure activities outside in nature, such as hiking, bird watching, swimming, skiing, other outdoor activities or just relaxing?} Answers on a 5-point scale: daily (coded 5), several times a week (4), several times a month (3), several times a year (2), and never (1);
    \item[MEAT] \emph{In a typical week, on how many days do you eat beef, lamb, or products that contain them?} Answers on a 8-point scale, 0 (coded 8), 1 (7), 2 (6), 3 (5), 4 (4), 5 (3), 6 (2), 7 (1), where numbers between brackets indicate our coding with higher numbers for more pro-environmental behaviour;  
    \item[RECYCLE] \emph{How often do you make a special effort to sort glass or tins or plastic or newspapers and so on for recycling?} Answers on a 4-point scale, always (4), often (3), sometimes (2), never (1);
    \item[AVOID] \emph{How often do you avoid buying certain products for environmental reasons?} Answers on a 4-point scale: always (coded 4), often (3), sometimes (2), and never (1).
\end{itemize}

We are interested in the relationship between environmental concern and environmental efficacy and these four behavioural items. Environmental concern (EC) was measured by the following statement: \emph{Generally speaking, how concerned are you about environmental issues}, where participants could respond on a 5-point scale ranging from "Not at all concerned" (1) till "Very concerned" (5). Environmental efficacy (EE) was measured by averaging responses to seven statements that each had a 5-point answer scale ranging from Agree Strongly (5) to Disagree Strongly (1). We also use gender, age, and number of years of education as predictor variables in the analysis. In psychological research, the relationship between attitudes and behaviour is often of interest. Therefore, our primary focus is on the relationship between environmental concern and efficacy and the four behavioural items, controlling for the other variables. We will analyse the data using both the dominance and proximity perspective and contrast the two analyses. 

In Table \ref{tab:fit_issp1}, we show the fit statistics of both analyses in dimensionalities 1, 2, and 3. The AIC and BIC disagree on the optimal dimensionality, that is, AIC points to the three dimensional solutions, while BIC points towards the two-dimensional models. We will focus on the more parsimonious two-dimensional solutions. Comparing the dominance (CLRRR) and proximity (CLRMDU) perspective, it can be seen that the fit statistics for the proximity model are a little better with lower deviance and AIC. 

\begin{figure}[t]
\begin{center}
\includegraphics[width = 1.1\textwidth]{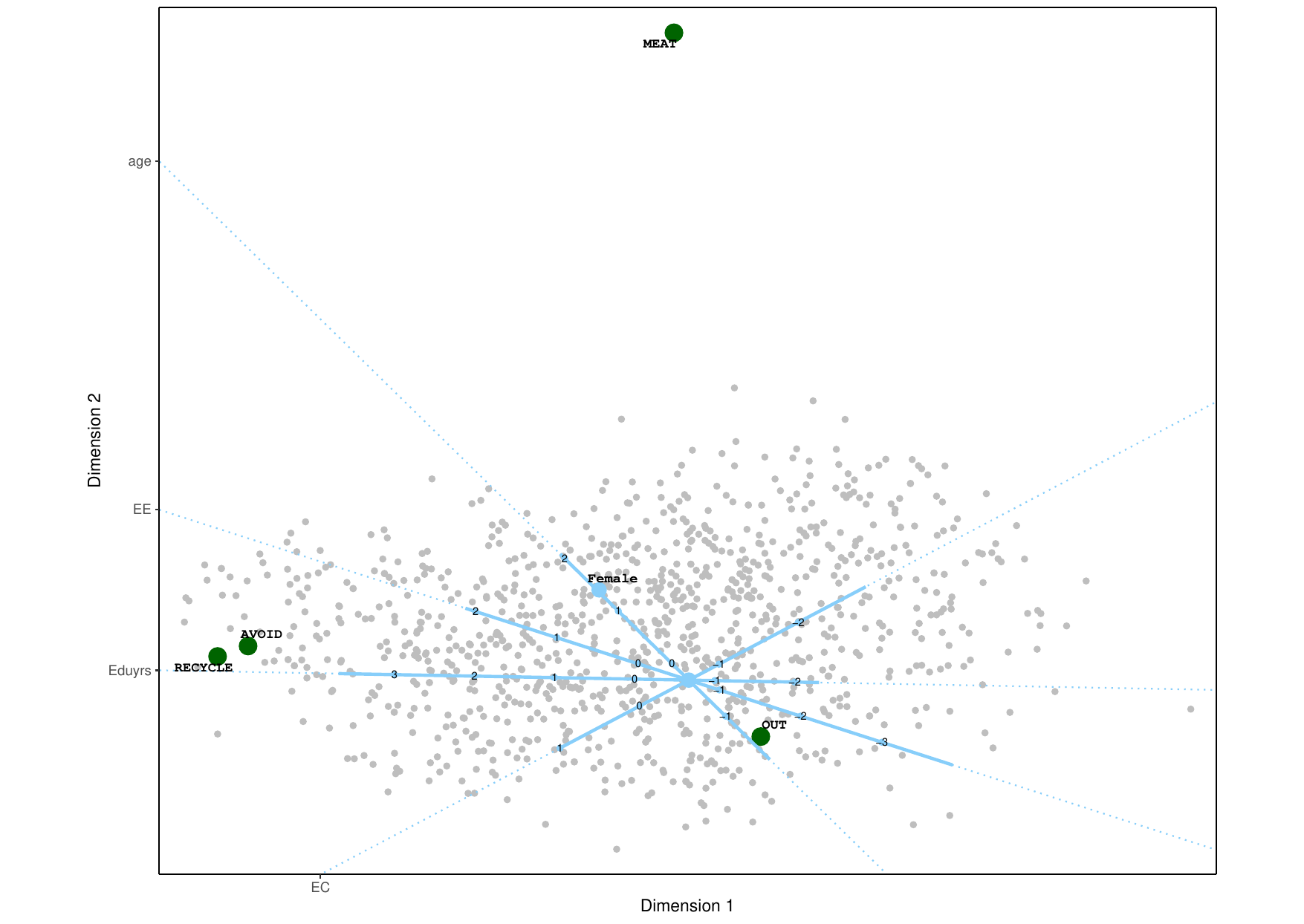}
\caption{Biplot for the cumulative logistic restricted multidimensional unfolding solution relating environmental attitudes with pro-environmental behaviour. }
\label{fig:issp_peb_clrmdu}
\end{center}
\end{figure}

We will inspect and interpret the visualization for the proximity model, the solution of the dominance model can be found in Appendix B (we will comment on it at the end of this section). The biplot is shown in Figure \ref{fig:issp_peb_clrmdu}, where it can be seen that the position of three of the four response variables is on the outside of the configuration, while one is more centrally located (i.e., OUT). This means that, while the predictors have a dominance relationship with the response variables MEAT, RECYCLE, and AVOID, they have a proximity relationship with OUT. 

\begin{table}[t]
\centering
\begin{tabular}{l|rrr|rrr}
  \hline
  & \multicolumn{3}{c|}{CLRRR} & \multicolumn{3}{c}{CLRMDU}\\
  dimensionality & deviance & AIC & BIC & deviance & AIC & BIC \\ 
  \hline
  1 & 11826.94 & 11876.94 & 12001.16 & 11839.25 & 11891.25 & 12020.44 \\ 
  2 & 11777.22 & 11839.22 & 11993.25 & 11756.83 & 11824.83 & 11993.77 \\ 
  3 & 11754.77 & 11824.77 & 11998.68 & 11733.53 & 11815.53 & 12019.25 \\ 
   \hline
\end{tabular}
\caption{Fit statistics for the dominance (CLRRR) and proximity (CLRMDU) analysis of pro-environmental behaviour in one, two, and three dimensions.}
\label{tab:fit_issp1}
\end{table}

Let us look at the biplot in more detail. The two predictor variables of interest, environmental concern (EC) and efficacy (EE), are represented by variable axes that run from the right hand side to the left hand side, meaning that persons that score high on these variables are located at the left hand side of the biplot, while persons that score low on these variables are positioned at the right hand side in the biplot. The control variables, education and age also run from right (lower scores) to left (higher scores). Females are located more to the top left compared to males. The higher a participant scores on EE or EC the closer they get to the response variables RECYCLE and AVOID, meaning that the higher a participant scores on the two attitude variables the more pro-environmental behaviour is reported. For these two variables only the radii for \emph{never}~|~\emph{sometimes} (1|2) and \emph{sometimes}~|~\emph{often} (2|3) are positive, meaning that participants are never classified in the response category \emph{always}. The biplots including the decision regions per response variable are shown in Figure B1 in Appendix B. These two response variables follow a dominance response process because all participant are located on one side of the point representing the response variable. 

The response variable OUT lies close to the origin of the two-dimensional solution. The radii for boundaries \emph{never}~|~\emph{several times per year} (1|2) and \emph{several times per year}~|~\emph{several times per month} (2|3) are positive. No participant will be classified in the class of \emph{daily} (5) or \emph{several times per week} (4) outside activities, as the estimated radius for these boundaries is negative. Note that younger participants are closer to the point for this response variable, indicating younger people more often engage in outside activities. With increasing scores on environmental efficacy (and average values for the other variables) the behaviour changes from \emph{several times per year}, to \emph{several times per month}, to \emph{several times per year}, and extrapolating to \emph{never}, a single peaked response pattern. A similar response function can be obtained for the predictor environmental concern (EC). 

Finally, we inspect the response variable MEAT. All radii are positive and therefore participants can be classified in each of the categories. The classification regions are shown in Figure B1, where for this data set the participants fall in 0, 1, 2, and 3 days per week eating meet (notice the reverse where not eating meat (0 days) is coded as 8 indicating the category with most pro-environmental behaviour). No participant is classified as eating meat more days per week, these regions fall on the lower bottom of the figure, where no participants are located. We see that some participants will be classified as typically eating no meat (within the smallest circle around MEAT). There is not a single predictor pointing in this direction, but a combination of female, higher age, relatively low EC, and a low number of years of education will result in this classification. 

As pointed out before, the answers to three out of the four response variables follow a dominance pattern, whereas only the responses to OUT follow a proximity pattern. This response variable (OUT) is probably the reason that the proximity model fits these data better. The biplot for the two-dimensional dominance model is shown as Figure B2, where the interpretation for MEAT, AVOID and RECYCLE closely follows the interpretation given here. The direction of the response variables is very similar to the positions in Figure \ref{fig:issp_peb_clrmdu}. The pattern for response variable OUT is different, now being monotonic and negatively related with environmental efficacy, age, and being female. 

\subsection{ISSP data: Environmental Efficacy}

In this second analysis using the ISSP data, we focus on differences between countries concerning environmental efficacy, controlling for gender, education and age. The seven items related to environmental efficacy are:
\begin{itemize}
    \item[1] It is just \underline{too difficult} for someone like me to do much about the environment;
    \item[2] I \underline{do} what is \underline{right} for the environment, even when it costs more money or takes more time;
    \item[3] There are \underline{more important} things to do in life than protect the environment;
    \item[4] There is no point in doing what I can for the environment unless \underline{others} do the same;
    \item[5] Many of the claims about environmental threats are \underline{exaggerated};
    \item[6] I find it \underline{hard to know} whether the way I live is helpful or harmful to the environment;
    \item[7] Environmental problems have a direct \underline{effect} on my everyday life.
\end{itemize}
Participants had to indicate on a five-point Likert scale for each of these statements whether they agreed strongly (coded as 5), agreed (4), are neutral (3),  disagreed (2), or disagreed strongly (coded as 1). 

As predictor variables, we use the 12 zero-one coded dummy variables for the countries, using Thailand as the control category (i.e., coded with 12 zeros), a dummy variable for gender where female is coded 1 and therefore male serves as a reference, and both the standardised scores of age and number of years of education as numeric predictors. The origin of the Euclidean space, therefore, corresponds to male participants from Thailand having average age and education. 

\begin{figure}[t]
\begin{center}
\includegraphics[width = .9\textwidth]{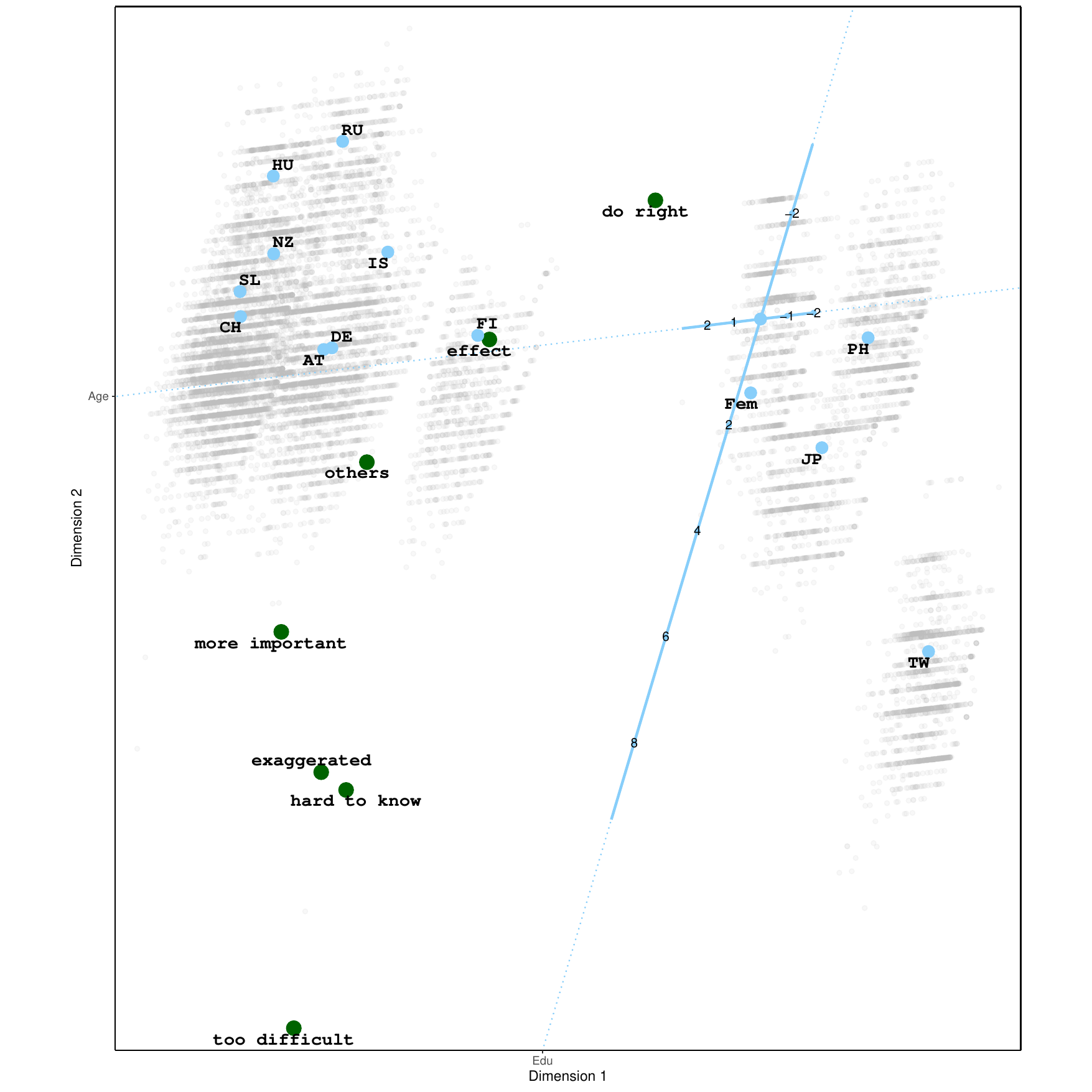}
\caption{Estimated configuration for Environmental Efficacy data.}
\label{fig:clmduissp}
\end{center}
\end{figure}

We fitted models in 1, 2, and 3 dimensions. The fit statistics are
\newline
\begin{table}[h]
\centering
\begin{tabular}{l|rrrr}
  \hline
  Dimensionality & deviance & npar & aic & bic \\ 
  \hline
  1 & 316958.57 & 50 & 317058.57 & 317444.02 \\ 
  2 & 313685.65 & 71 & 313827.65 & 314374.99 \\ 
  3 & 312597.80 & 91 & 312779.80 & 313481.32 \\ 
   \hline
\end{tabular}
\end{table}\newline
showing that the three-dimensional model fits best. For illustrative purposes, the two-dimensional biplot is shown in Figure \ref{fig:clmduissp}. The response variables are located from the bottom left (item 1, \underline{too difficult}) to top middle (item 2, \underline{do right}). They lie almost on a curve, where items 5 and 6 (\underline{exaggerated} and \underline{hard to know}) lie close together, indicating similar response tendencies for those two items. 

Looking at the predictor side we first note that the origin, where the variable axes for \emph{Education} and \emph{Age} cross represents male participants from Thailand with average age and education. We can see that \emph{Education} has a large influence on the positioning of the participants, where participants with a few years of education are in the top of the biplot and those with many years of education in the bottom. The variable axis for \emph{Age} is much shorter, and therefore \emph{Age} has a smaller effect on the outcomes. The last control variable is \emph{Gender}, where we see that the category Female is below the origin. Therefore, comparing female and male participants with the same values for the other predictor variables, the females are located below male participants and therefore closer to the items at the bottom of the configuration. The countries partition in two clusters, with on the right hand side of the biplot the Asian countries Japan, the Philippines, Taiwan and Thailand and on the left hand side a cluster of Hungary, Russia, New Zealand, Island, Slovenia, Switzerland, Austria, Germany, and Finland. Austria and Germany are very close, meaning that responses in these two countries follow a similar pattern. Russia and Hungary are close together and positioned more in the top of the biplot, further away from the response variables \underline{more important}, \underline{exaggerated}, \underline{hard to know}, and \underline{too difficult}, indicating lower probabilities of agreement with these items compared to  participants from other countries. 

\begin{figure}[t]
    \centering
    \begin{minipage}{.5\textwidth}
        \centering
        (a)\newline
        \includegraphics[width=0.9\textwidth]{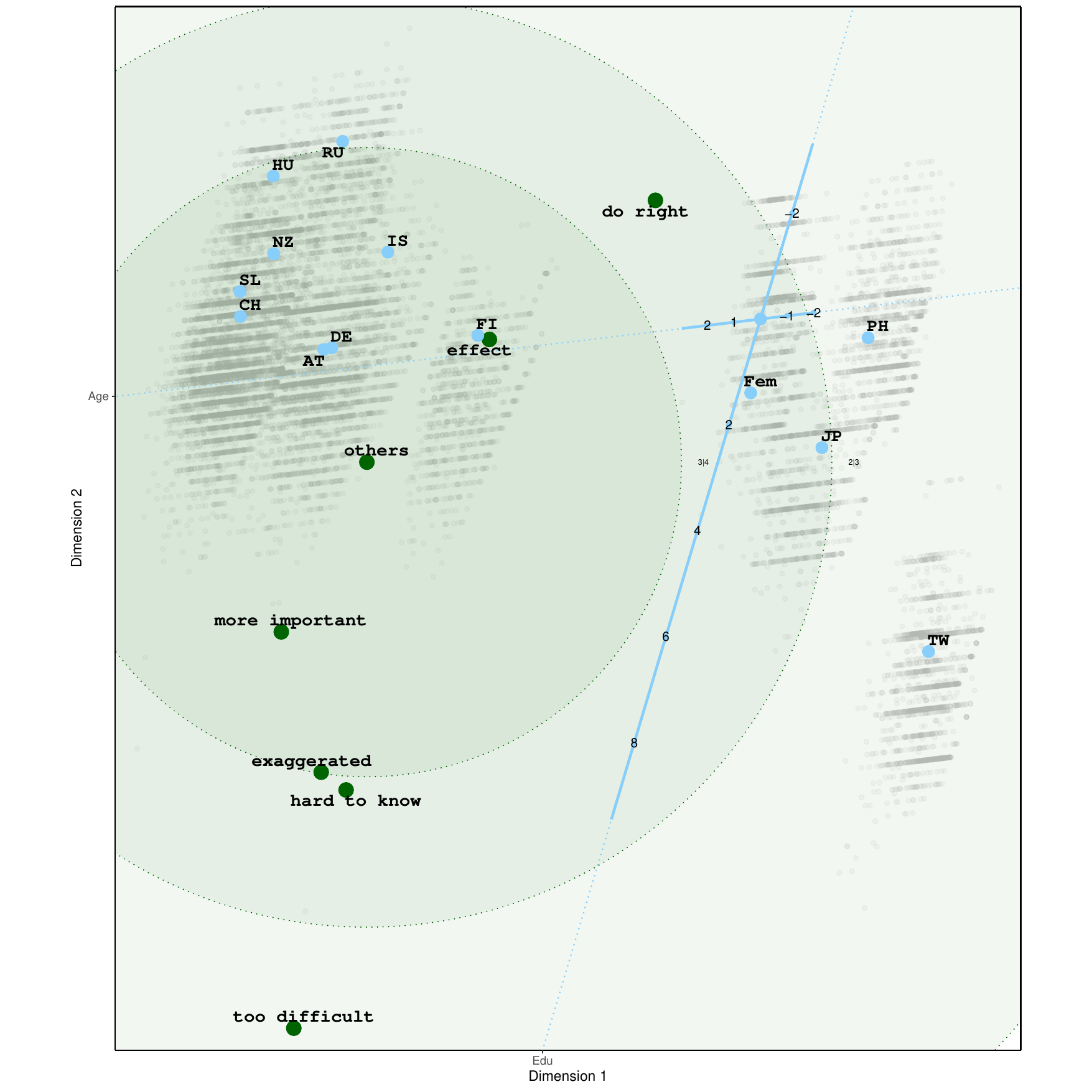}
    \end{minipage}%
    \begin{minipage}{0.5\textwidth}
        \centering
        (b)\newline
        \includegraphics[width=0.9\textwidth]{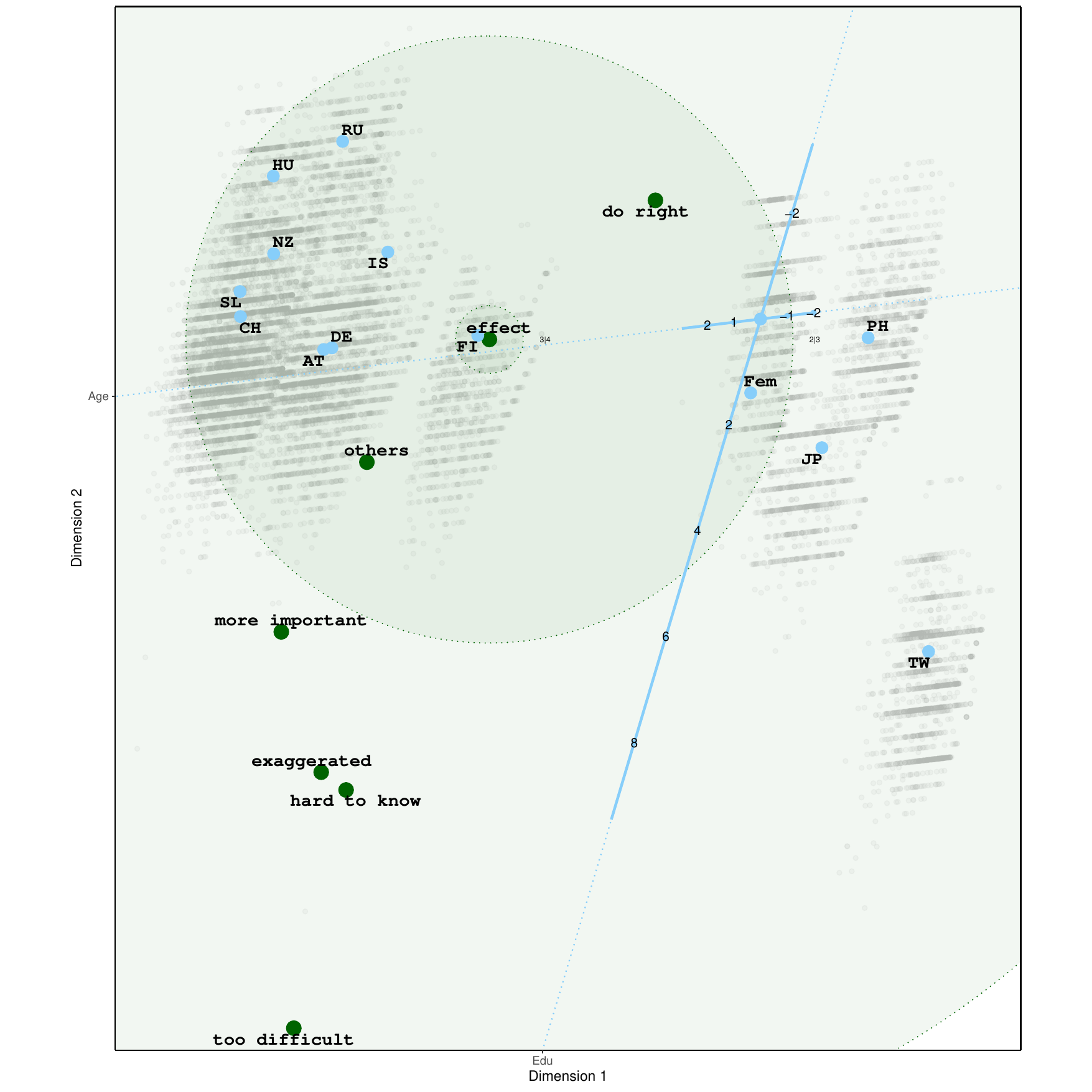}
    \end{minipage}
    \caption{Estimated configuration for Environmental Efficacy data with decision regions for Item 4 (a) and Item 7 (b).}
    \label{fig:clmduissp2}
\end{figure}

Now, let us look in more detail at two biplots where we included the circles representing the classification regions for items 4 (``There is no point in doing what I can for the environment unless \underline{others} do the same'') and 7 (``Environmental problems have a direct \underline{effect} on my everyday life''). The biplots are shown in Figure \ref{fig:clmduissp2}. For item 4, we see that the cluster of Asian countries on the left disagrees (all participants from Taiwan and most participants from the Philippines) or is neutral (most participants from Thailand and the higher educated and elderly in Japan). 

Participants from the other countries mainly agree with this statement, because their position is within the \emph{neutral}~|~\emph{agree} (i.e., 3|4) circle. An exception are the participants with only few years of education from Russia and Hungary, who tend to be neutral. 

Inspecting the biplot with classification regions for item 7, we can conclude that only participants from Finland with average age and education tend to agree with this item. Most participants from the Asian countries tend to disagree (except participants from Thailand), while most participants from the other countries tend to be neutral towards this item. It seems that especially \emph{Age} has an influence on this response variable as elderly people from the Asian countries have larger probabilities to agree, while the younger people from the other cluster have higher probabilities to agree (i.e., they are closer to the position of the item). 

\section{Simulation studies}\label{sec:simulations}

We conducted a simulation study to verify whether the algorithms work properly. Therefore, we use the estimated parameters from the second example, described in Section \ref{sec:peb}, specifically the matrices $\bm{B}$ and $\bm{V}$. The precise values are shown in Appendix C.  
In the data generation process, we start drawing predictor variables from the multivariate normal distribution with mean zero and covariance matrix equal to the correlation matrix of the predictors in Section \ref{sec:peb} (see Appendix C for values). The predictor variables and the population matrix $\bm{B}$ defines $\bm{U}$. With these coordinates and the matrix $\bm{V}$, values on the latent variables (i.e., the $\theta_{ir}$'s) can be computed. For the dominance models we use the inner product ($\theta_{ir} = \langle \bm{B}'\bm{x}_i, \bm{v}_r \rangle$), while for the proximity models we use the distances ($\theta_{ir}  = -d(\bm{B}'\bm{x}_i, \bm{v}_r)$). With these values and the threshold parameters, probabilities of the response categories for each of the response variables can be obtained. We draw observed outcome variables for each response variable independently from the multinomial distribution. 

We vary sample size with values 250, 500, and 1000. We also vary whether the response variables have three or five categories and we manipulate the number of response variables, to be equal to 4 or 8. Note that in the empirical example we had 4 response variables. We created 4 others by simply rotating $\bm{V}$ by $45^o$ and adding the obtained coordinates to the matrix (see Appendix C for population values). We used a full factorial $3 \times 2 \times 2$ design, with 200 replications per condition.

The simulation is done separately for the cumulative logistic reduced rank model and the cumulative logistic restricted multidimensional unfolding. 

As outcome variable we take the following measure of recovery: 
\[
\delta = \sqrt{\frac{\sum_i \sum_r (\theta_{ir} - \hat{\theta}_{ir})^2}
{\sum_i \sum_r \hat{\theta}^2_{ir}}},
\]
where $\theta_{ir} = \langle \bm{B}'\bm{x}_i, \bm{v}_r \rangle$ for the dominance model and $\theta_{ir}  = -d(\bm{B}'\bm{x}_i, \bm{v}_r)$ for the proximity model.
This measure resembles the Stress-1 value that is often used in multidimensional scaling and unfolding. The benefit of this measure over, for example, looking at the recovery of $\bm{B}$ or $\bm{V}$ is that we do not need to take into account any reflection, rotation, or other transformations to overcome indeterminacies. We like to point out that the measure $\delta$ is not necessarily comparable between the dominance and proximity perspective as for the proximity models the values of $\theta_{ir}$ are all negative, while for the dominance models they can be both positive and negative.

\begin{figure}[ht!]
\begin{center}
\includegraphics[width = .9\textwidth]{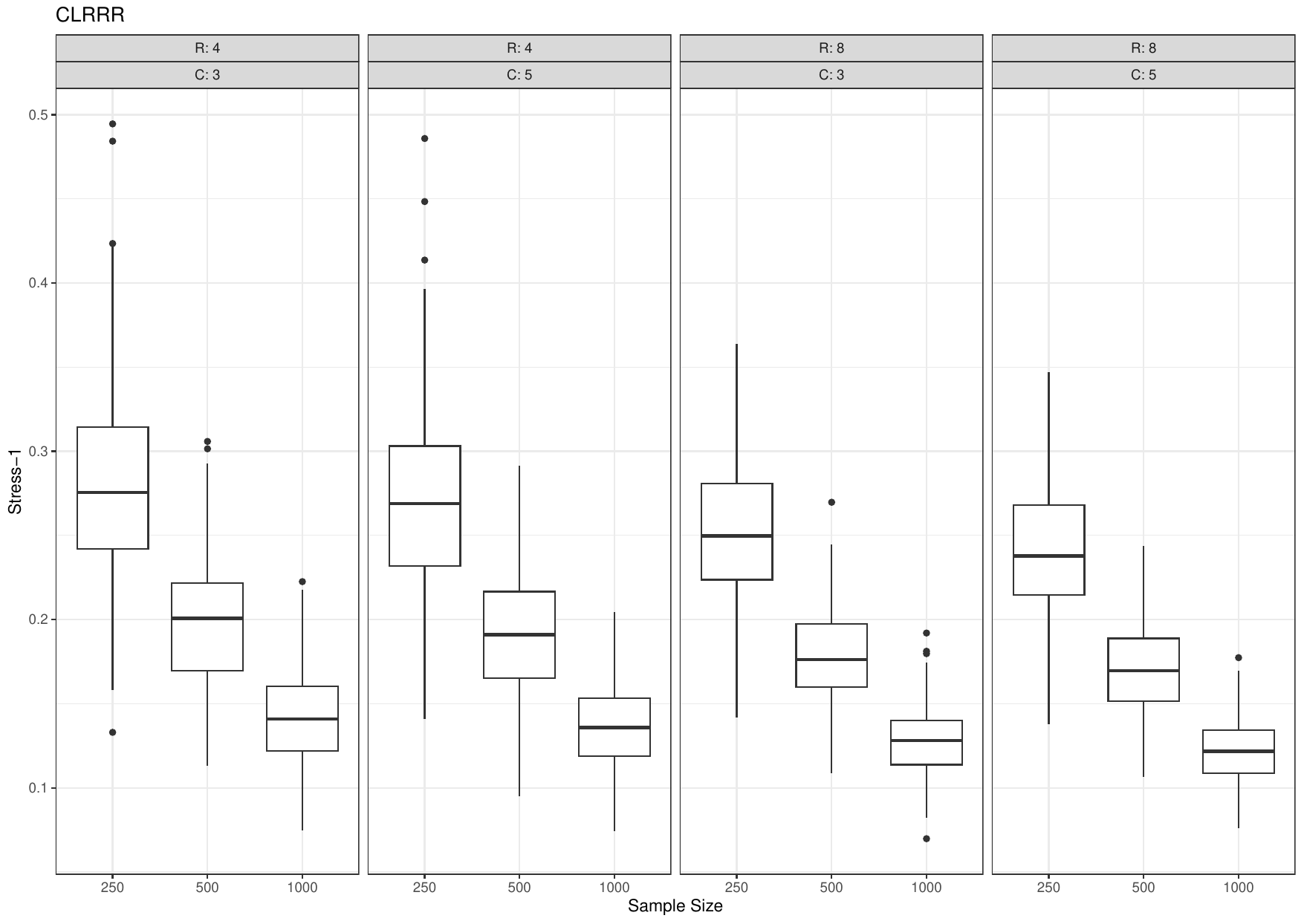}
\caption{Simulation results for cumulative logistic reduced rank regression. $R$ denotes the number of response variables, $C$ the number of response categories per response variable. On the horizontal axis we show the different sample sizes, while on the vertical axes, the value of recovery is found where lower values represent better recovery.}
\label{fig:simclrrr}
\end{center}
\end{figure}

\begin{figure}
\begin{center}
\includegraphics[width = .9\textwidth]{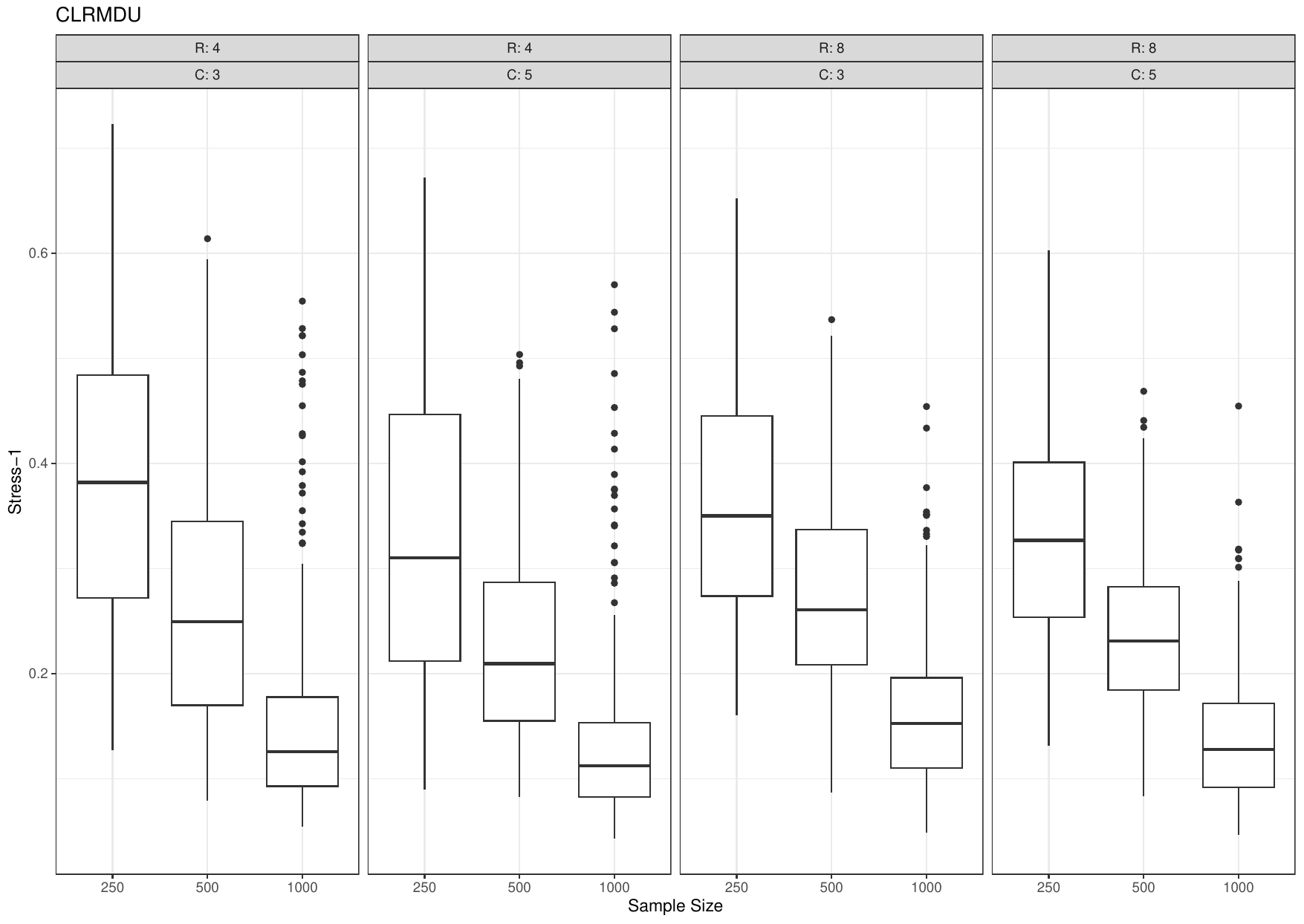}
\caption{Simulation results for cumulative logistic restricted multidimensional unfolding. $R$ denotes the number of response variables, $C$ the number of response categories per response variable. On the horizontal axis we show the different sample sizes, while on the vertical axes, the value of recovery is found where lower values represent better recovery.}
\label{fig:simclrmdu}
\end{center}
\end{figure}

The results of our simulation studies are presented in Figures 
\ref{fig:simclrrr} for the reduced rank model and Figure \ref{fig:simclrmdu} for the restricted multidimensional unfolding. In both cases, we see that recovery is good and that it improves with increasing sample size, number of response variables ($R$), and number of response categories ($C$). The changes for the number of response variables and the number of response categories are small, but the values of $\delta$ become smaller and less variable. Sample size has a larger influence as is clear from the boxplots.

\section{Discussion}

In this manuscript, we proposed a novel framework for multidimensional analysis of ordinal data in a cumulative logistic framework. Logistic regression models are the usual models of choice when the response variable are categorical. For an ordinal response variable, the cumulative logistic regression model, also known as the proportional odds regression model, is a typical analysis model often used in practice. We developed cumulative logistic models for multiple ordinal response variables that we like to analyze together.  

We distinguished between models for proximity and dominance items. For proximity items, distance representations are used, whereas for dominance items inner product representations are used. A distance representation is obtained using multidimensional unfolding, whereas an inner product representation is obtained using principal components. When predictor variables are available for the participants, restricted versions of both types of models are obtained, leading to reduced rank regression and restricted multidimensional unfolding. 

For maximum likelihood estimation, an Expectation Majorization Minimization (EMM) algorithm was developed and tested. In the E-step the conditional expectation of the complete data negative log likelihood is taken which is majorized by a least squares function. This least squares function is minimized using well known updating steps. For the estimation of the threshold, standard maximum likelihood methods can be employed, using the structural part as an offset. The algorithm monotonically decreases the negative log likelihood function. In a simulation study, we tested the performance of the algorithm and we showed that overall the recovery is good. In more detail, the recovery increases with sample size, more response variables, and increasing number of response categories. 

Biplot representations were discussed in detail. For cumulative logistic PCA similar biplots have been proposed before by \cite{vicente2014logistic}. Our CLPCA biplots differ in the way the markers for the response variable axes are defined. We use the underlying latent variable, whereas \cite{vicente2014logistic} use posterior probabilities. The advantage of our idea is that all subsequent markers are represented, whereas the posterior probability of some categories of an outcome variable might never be higher than the probabilities of the rest. Furthermore, our markers directly follow from the model estimates, whereas the posterior probabilities need an extra computational step. The CLMDU biplots are, as far as we know, new. In the CLMDU biplots, the distances between points for the participants and points for the response variables determine the probabilities. The threshold parameters can be included in the biplot as circles. For a response variable with $C_r$ categories,  $C_r -1$ circles are drawn partitioning the space in $C_r$ regions. 

We showed applications of the analysis framework for three empirical data sets. The response variables in the first data set are ten exam items about the application of statistics. These items are typical cognitive response variable and therefore are of the dominance type. Besides the responses to ten items also characteristics of the students are available. We used these characteristics as predictor variables. For the analysis of these data, we employed the cumulative logistic reduced rank model and show the results both in terms of a table with regression coefficients (Table A2) and a biplot (Figure \ref{fig:student}). Each of the columns of the coefficient matrix can be interpreted in a similar way as a standard proportional odds model. Although this allows for a detailed interpretation of the effect of each predictor on each response, it is difficult to obtain a holistic interpretation of the model or a view about the dependencies among the response variables from this table. The biplot enhances these two latter aspects. We saw that a group of eight response variables are in a similar way related to the predictor variables, the variable axes for these response variables all run from left to right. Two response variables behave differently, running from the bottom to the top (A2g and A1d). We can infer that there are strong dependencies among the group of eight and the group of 2. Furthermore, we can obtain a holistic view on the model by looking at the angles of the predictor and response variable axes. As there is a group of predictor variables with axes running from top left to bottom right (APS, SATSv, SATSc, ENGc) and another from left to right (SATSa, PMP, ENGb) we can easily remember that those two groups act differently on the response variables. The variables within either of the two groups act similarly on the response variables. Specific effects can be obtained from the table again. We see that the table and biplot reinforce each other in interpretation.  

In the second data set there are four behavioural response items. Predictor variables include gender, age, education, and environmental efficacy and concern. We used this data set to contrast the distance and inner product approach. The data were analysed by both approaches and it was shown that the distance approach fits the data better, mainly because of one of the response variables. In the biplot visualization, the other three response variables are represented by a point in the periphery of the plot indicating that a dominance perspective suits these three items well. The response variable OUT, however, acts like a proximity item: participants answering that they never engage in outdoor activity have different characteristics in terms of age, number of years of education, environmental concern and environmental efficacy. 

Using the third data set, we applied the distance approach. The response variables measure attitudes which usually are considered proximity variables. Given age, gender, and education, the goal was to see whether there are differences between countries. We showed the resulting biplot visualization and gave a detailed interpretation indicating the differences between countries and the influence of education, age and gender. The application of our framework to these data sets shows the versatility of our approach. 

The framework we proposed gives the opportunity to analyse complex data sets having ordinal response variables with dimension reduction techniques. The response variables do not need to be indicators of underlying constructs as in psychological scales but simply may be a number of related questions on a given topic (like pro environmental behaviour). The framework is applicable even in the case of a limited number of response variables, in contrast to item response models that usually require many items (see below for a further comparison). The models for proximity items and the corresponding visualizations with biplots using circles as classification regions are new and are a valuable tool for data analysis. Biplots for ordinal response variables have been proposed by \citep{vicente2014logistic}. They, however, only show methods for dominance variables without predictor variables. We extended their method by constraining the principal scores using predictor variables (i.e., the reduced rank regression approach). We also gave an alternative way for adding markers to the variable axes, one that is in line with the underlying variable idea. 

In our multidimensional models a proportional odds assumption is made, as discussed in Section 2.1. It is not easy to test the validity of this assumption within our modelling approach. One approach would be to fit different models that do not make this assumption and compare the fit of the two models. For the CLPCA model, we could compare against multinomial multiple correspondence analysis \citep[MMCA; ][]{groenen2016multinomial}. For CLRRR we could compare to a constraint version of MMCA (which first needs to be developed). Such comparisons are related to the score test for the proportional odds assumption \citep[][Section 7.2]{agresti2002categorical}. This score test tests whether the effects are the same for each cumulative logit against the alternative of separate effects. 
If the fit of the MMCA models is much better this is an indication of a violation of the proportional odds assumption for one or more response variables. For which response variable the assumption is violated then needs further investigation. An alternative applicable for CLRRR, is to apply the score test for each response variable separately when fitting $R$ different models. When none of these tests suggest a rejection of the null hypothesis, we can safely conclude that the assumption is also valid for the multivariate model. When one of the tests rejects the null hypothesis, we directly know for which response variable the proportional odds assumption is false. For the distance models (CLMDU and CLRMDU) validating the proportional odds assumption is more difficult. We could, for instance, develop new models that combine the cumulative logit models developed here with multinomial restricted unfolding \citep{derooijbusing2022}, a distance model for nominal variables, and investigate whether such a new model signals violations of the assumption. Further work is needed to find ways to verify this assumption within the framework presented. 

On the other hand, \cite{harrell2001regression} points out that we should not worry too much about the proportional odds assumption. Rank order based statistics such as the Wilcoxon test and the Kruskal-Wallis tests are special cases of the proportional odds regression model. Furthermore, rank order correlations are closely related to the proportional odds model. It seems best, to place our models in a bias-variance trade-off perspective. The proportional odds assumption might not be valid for some response variables leading to biased results. However, adding parameters to avoid the assumption might lead to more variance and, when the extra variance exceeds the bias, lead to worse model performance.   

Sometimes, a priori information is available about which response variables group together on specific dimensions. This would entail that elements of the matrix $\bm{V}$ are set to zero. Similar constraints could, in theory, be imposed on the matrix $\bm{B}$, specifying that some predictor variables are connected to specific dimensions. At the moment it is not possible to use such information in the analysis. Further research is needed to incorporate such knowledge. 

Logistic multidimensional models are closely related to Item Response Models. In fact, the model formulation of cumulative logistic PCA in one dimension is equivalent to Samejima's Graded Response Model \citep[GRM;][]{samejima1969estimation}. In the GRM, however, the principal scores are usually assumed to be a random effects from  a normal distribution. In PCA, principal scores are fixed effects. Also explanatory versions of the GRM have been proposed \citep{tuerlinckx2004models} to include predictor variables. Item response models are usually targeted towards test construction. Our models focus more on dimension reduction and prediction of correlated response variables. We like to note that these graded response models also make the proportional odds assumption, similar to our approaches, although not much emphasis is placed on this assumption in the literature. Within the context or item response theory, also single peaked items can be modelled. The best known model for ordinal data is the generalized graded unfolding model \citep[GGUM;][]{roberts2000general}. The model definition is quite involved, defined by two GRMs, one from below and one from above. The GGUM model is a unidimensional model, no multidimensional generalizations have been proposed so far, although recently a R-package for estimation of such multidimensional models has been proposed \citep{tu2021bmggum}.  Recently some explanatory versions, including predictors for the observations have been proposed by \cite{usami2011generalized} and \cite{joo2022explanatory}. Again, those item response models are targeted towards test construction, whereas our approaches are targeted towards dimension reduction and prediction of correlated response variables. Maximum likelihood estimation of logistic models with normally distributed random effects is generally problematic due to an intractable integral \citep{tuerlinckx2006statistical}. In our logistic multidimensional data analysis techniques, we do not have those random effects, so maximum likelihood estimation is simpler. The goals of the two approaches are different. Item response models usually are targeted towards optimal latent trait estimation. Often, a priori knowledge is available on the traits under investigation. External information (i.e., predictor variables) is used to address sub population heterogeneity or to increase estimation accuracy. The goals of our analysis framework is more towards dimension reduction to obtain insight into the structure of the response variables or, when predictor variables are available, to develop simultaneous regression models for the response variables in a reduced dimensional space. 

In our framework, we focused on predictor variables describing the participants. In some situations, external information about the items might be available. In our analysis, we could linearly constrain the matrices $\bm{V}$ to include such information. For the type of response variables considered in our applications, however, typically no further information is available. Further research and programming is needed to incorporate such constraints. 

In conclusion, we proposed a family of models for multidimensional analysis of multiple ordinal response variables. The framework contains four different models. We distinguished between models for dominance variables and proximity variables. Within each we distinguished models with and without predictor variables. Algorithms for all methods proposed in this paper are implemented in the R software. The logistic mapping package \citep{lmappackage} contains the functions \texttt{clpca} and \texttt{clmdu} and corresponding plotting function that can be used for the analysis described in this paper. 

\pagebreak
\section*{Statements and Declarations}

\paragraph*{Thanks:} 

The authors would like to thank Dr. Rosa Fabbricatore for making the data available to us. We would also like to thank the three anonymous reviewers for their helpful input. 

The first author revised the manuscript while he was a fellow at the Netherlands Institute of Advanced Studies in Amsterdam. 

\paragraph*{Data availability:} 

The data used in this study is available from GESIS \citep{ISSP2020data}, see \url{https://www.gesis.org/en/issp/modules/issp-modules-by-topic/environment/2020}

The data set about Students performance on the statistics exam can be requested from Dr. Rosa Fabbricatore

\paragraph*{Funding:}
The authors declare that no funds, grants, or other support were received during the preparation of this manuscript.

\paragraph*{Competing Interests:}
The authors have no relevant financial or non-financial interests to disclose.

\paragraph*{Author Contributions:}
MdR initiated and conceptualized this study. MdR developed the algorithm and implemented it in R. FB wrote C-code for multidimensional unfolding and restricted multidimensional unfolding. DW tested the CLMDU and CLRMDU algorithms and applied them to the ISSP data. DW developed the visualization of MDU and RMDU analysis under the supervision of MdR. LB tested the CLPCA and CLRRR algorithms and applied them to the ISSP data.
MdR wrote the first draft of the paper. FB, LB, and DW commented on the first draft. MdR revised the manuscript on the basis of th reviewers comments. All authors approved the final manuscript. 

\clearpage

\section*{Appendix A: Statistics for Student data}

We use different labels for the response variables than the authors in \cite{fabbricatore2024} use. Here is a table relating our labels (left column) to the names shown in their paper (center column) and the actual question for the students:
\begin{table}[h]
\centering
\begin{tabular}{llp{8cm}}
  \hline
  A1a & T1\_ClassVar\_A & Select the type of variable that best describes the gross annual income\\ 
  A1b & T1\_GraphForQuant\_A & Find the error in a graph representing the distribution of a continuous variable (i.e.  waiting time in minutes)\\ 
  A1c & T1\_Median\_A & Calculate the median of individual series data\\ 
  A1d & T1\_ArithmeticMean\_A & Calculate the overall weighted arithmetic mean given the mean of three groups of individuals with different class sizes\\ 
  A1g & T1\_SkewnessNKurtosis\_A & Calculate the Hotelling-Solomon skewness coefficient\\ 
  A2a & T2\_SamplingSpace\_A & Calculate the size of the sample space if 1 coin and 2 six-sided dice are tossed together at the same time\\ 
  A2c & T2\_ConditionalProb\_A & Calculate the conditional probability, given the marginal probabilities and the joint probability\\ 
  A2d & T2\_Bernoulli\_A & Calculate the variance of a Bernoulli trial given the probability of success\\ 
  A2g & T2\_Gaussian\_A & Calculate probability for normally distributed data\\ 
  A2i & T2\_SamplingVariance\_A & Determine the sample size using the formula derived from the corrected sample variance\\ 
   \hline
\end{tabular}
\end{table}

The model selection procedure is summarized using Table A1. 
\begin{table}[ht]
\centering
\begin{tabular}{lrrrr}
  \hline
    & npar & deviance & aic & bic \\ \hline
  1 & 48.00 & 2310.47 & 2406.47 & 2546.98 \\ 
  2 & 74.00 & 2250.50 & 2398.50 & 2615.12 \\ 
  3 & 98.00 & 2206.37 & 2402.37 & 2689.24 \\ \hline
  Age & 72.00 & 2251.38 & 2395.38 & 2606.14 \\ 
  Gender & 72.00 & 2255.71 & 2399.71 & 2610.48 \\ 
  Math & 72.00 & 2277.05 & 2421.05 & 2631.82 \\ 
  Statistical.Anxiety & 68.00 & 2253.18 & 2389.18 & 2588.23 \\ 
  SATS & 66.00 & 2271.74 & 2403.74 & 2596.94 \\ 
  MSLQ & 66.00 & 2257.26 & 2389.26 & 2582.46 \\ 
  APS & 72.00 & 2258.28 & 2402.28 & 2613.04 \\ 
  AMS & 72.00 & 2250.55 & 2394.55 & 2605.31 \\ 
  ENG & 68.00 & 2265.35 & 2401.35 & 2600.40 \\ \hline
\end{tabular}
\caption*{Table A1: Fit statistics for the student data. First three rows show fit statistics of models in 1 to 3 dimensions including all predictors. The remaining rows show fit statistics leaving out one variable or (set of) scales. \texttt{npar} denotes the number of parameters.}
\end{table}

The implied coefficients ($\bm{BV}'$) are shown in Table A2, and can simply be understood similarly as coefficients in a cumulative logistic model (proportional odds model). 

\begin{table}[ht]
\centering
\begin{tabular}{lrrrrrrrrrr}
  \hline
 & A1a & A1b & A1c & A1d & A1g & A2a & A2c & A2d & A2g & A2i \\ 
  \hline
0$|$1 & -0.85 & -2.98 & -4.03 & -1.38 & -2.53 & -2.57 & -0.88 & -2.03 & -1.26 & -0.92 \\ 
  1$|$2 & 1.22 & 0.30 & -0.45 & -0.34 & -1.78 & -2.07 & 0.42 & -1.79 & 0.06 & 1.18 \\ \hline
  G & -0.03 & -0.61 & -0.19 & -0.00 & -0.17 & -0.68 & -0.30 & -0.61 & 0.27 & -0.96 \\ 
  PMP & 0.05 & 0.37 & 0.29 & 0.38 & 0.23 & 0.55 & 0.27 & 0.54 & 0.45 & 0.47 \\ 
  SATSa & 0.02 & 0.11 & 0.14 & 0.23 & 0.10 & 0.20 & 0.11 & 0.21 & 0.31 & 0.11 \\ 
  SATSc & 0.01 & -0.26 & 0.05 & 0.28 & 0.02 & -0.20 & -0.06 & -0.14 & 0.56 & -0.49 \\ 
  SATSv & 0.01 & -0.20 & 0.02 & 0.17 & -0.00 & -0.17 & -0.06 & -0.13 & 0.36 & -0.37 \\ 
  SATSd & -0.02 & -0.06 & -0.13 & -0.23 & -0.09 & -0.14 & -0.08 & -0.16 & -0.34 & -0.02 \\ 
  APS & 0.00 & -0.25 & 0.00 & 0.17 & -0.02 & -0.23 & -0.09 & -0.18 & 0.38 & -0.45 \\ 
  ENGa & 0.01 & 0.02 & 0.07 & 0.13 & 0.05 & 0.07 & 0.04 & 0.08 & 0.20 & -0.00 \\ 
  ENGb & 0.00 & 0.14 & 0.03 & -0.02 & 0.03 & 0.15 & 0.06 & 0.13 & -0.09 & 0.23 \\ 
  ENGc & -0.01 & 0.21 & -0.06 & -0.26 & -0.02 & 0.15 & 0.05 & 0.10 & -0.51 & 0.41 \\ 
   \hline
\end{tabular}
\caption*{Table A2: Implied coefficient for the student data}
\end{table}

\clearpage

\section*{Appendix B: Pro-environment behaviour}

The biplots for the cumulative logistic restricted multidimensional unfolding analysis including the circles for each of the response variables is shown in Figure B1. 

\begin{figure}[t]
\begin{center}
\includegraphics[width = 1.1\textwidth]{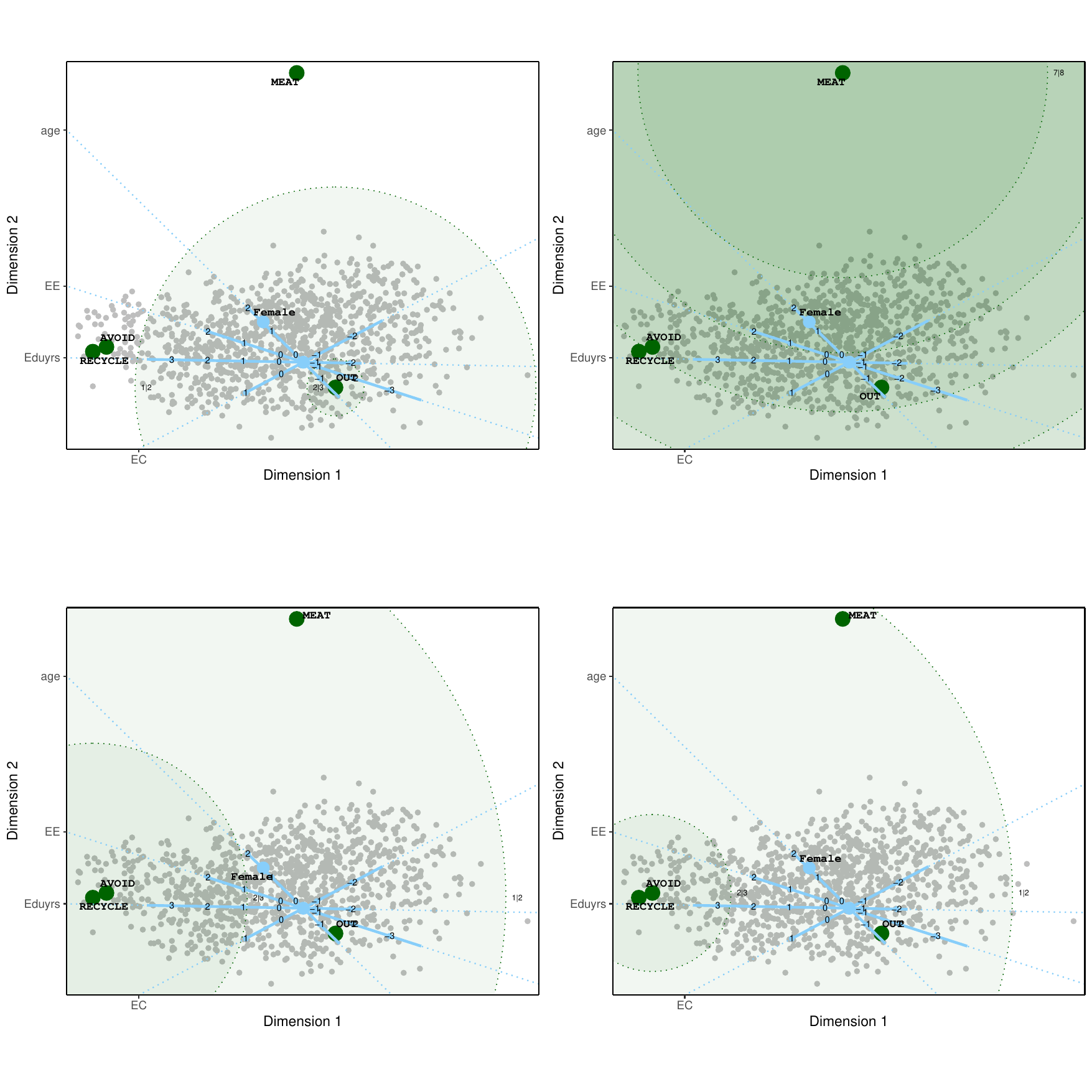}
\caption*{Figure B1: Biplot for the cumulative logistic restricted multidimensinal unfolding solution relating environmental attitudes with pro-environmental behaviour. }
\label{fig:issp_peb_clrmdu_all}
\end{center}
\end{figure}

The biplot for the cumulative logistic reduced rank model is shown in Figure B2. 

\begin{figure}[t]
\begin{center}
\includegraphics[width = 1.1\textwidth]{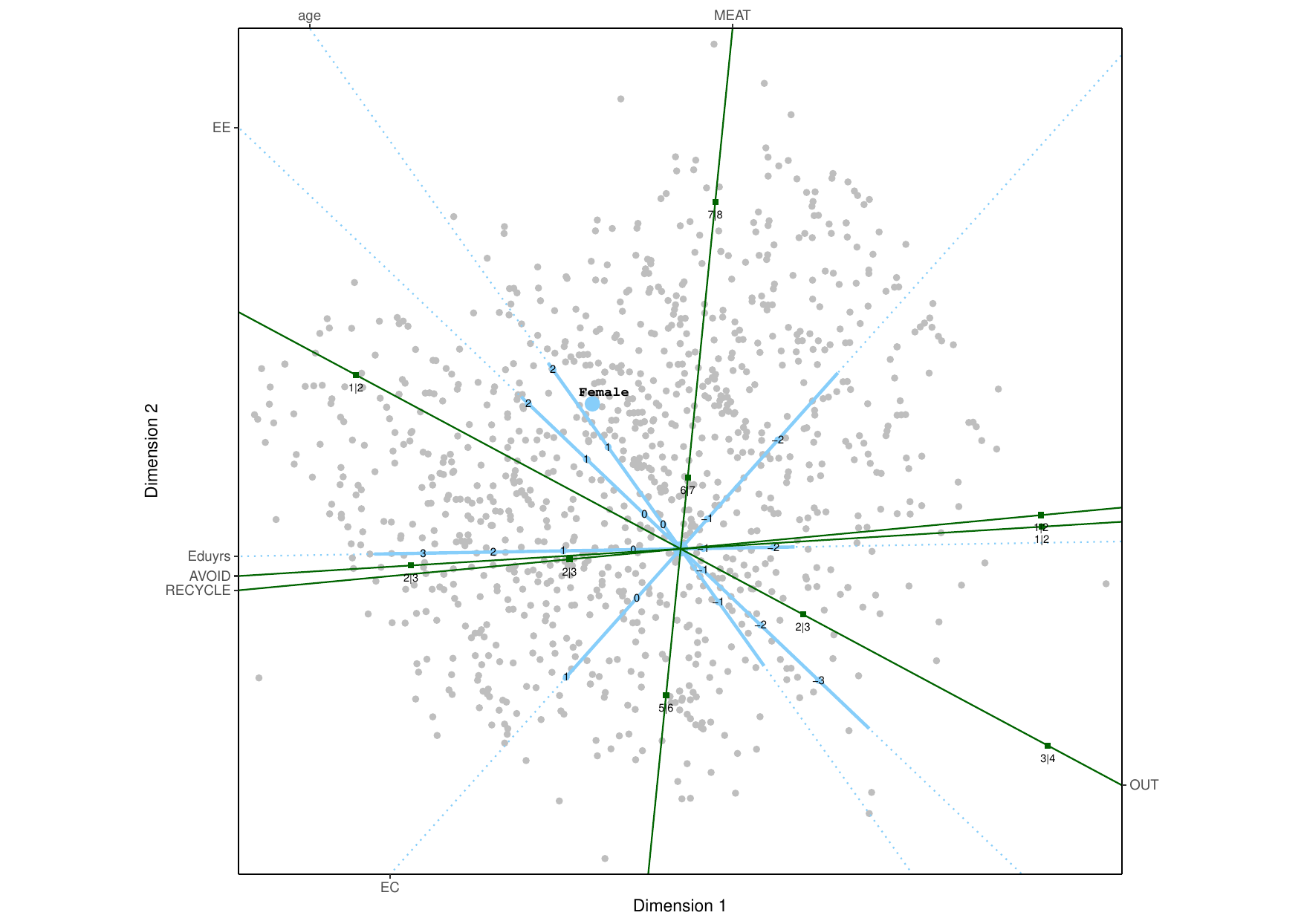}
\caption*{Figure B2: Biplot for the cumulative logistic reduced rank model relating environmental attitudes with pro-environmental behaviour. }
\label{fig:issp_peb_clrrr}
\end{center}
\end{figure}

\clearpage

\section*{Appendix C: Population parameters for Simulation studies}

The predictor variables are sampled from a multivariate normal distribution with zero means and covariance matrix 
\begin{table}[h!]
\centering
\begin{tabular}{lrrrrr}
  \hline
 & X1 & X2 & X3 & X4 & X5 \\ 
  \hline
  X1 & 1.00 & 0.01 & -0.02 & 0.01 & 0.04 \\ 
  X2 & 0.01 & 1.00 & -0.59 & 0.19 & 0.16 \\ 
  X3 & -0.02 & -0.59 & 1.00 & -0.00 & -0.00 \\ 
  X4 & 0.01 & 0.19 & -0.00 & 1.00 & 0.25 \\ 
  X5 & 0.04 & 0.16 & -0.00 & 0.25 & 1.00 \\ 
   \hline
\end{tabular}
\end{table}

The matrix with population coefficients $\bm{B}$ equals the estimated coefficients from the example in Section 4.2, that is
\begin{table}[h!]
\centering
\begin{tabular}{rrr}
  \hline
 & 1 & 2 \\ 
  \hline
  X1 & -0.16 & 0.19 \\ 
  X2 & -0.37 & 0.04 \\ 
  X3 & -0.17 & 0.19 \\ 
  X4 & -0.40 & -0.17 \\ 
  X5 & -0.28 & 0.12 \\ 
   \hline
\end{tabular}
\end{table}

The population matrix $\bm{V}$ equals

\begin{table}[h]
\centering
\begin{tabular}{rrr}
  \hline
 & 1 & 2 \\ 
  \hline
  Y1 & 0.44 & -0.45 \\ 
  Y2 & 0.34 & 2.35 \\ 
  Y3 & -1.68 & 0.05 \\ 
  Y4 & -1.55 & 0.08 \\ 
  Y5 & -0.16 & -0.61 \\ 
  Y6 & 2.18 & 0.94 \\ 
  Y7 & -0.84 & 1.45 \\ 
  Y8 & -0.74 & 1.36 \\ 
   \hline
\end{tabular}
\end{table}

where we use only the first four rows when $R =4$ and the complete matrix when $R = 8$

The parameters $\bm{m}$ are equal for the different response variables. The threshold values in case of three categories are -1.0 and -0.5, while the values with five response categories are -2.0, -1.5, -1.0, and -0.5.

\clearpage

\bibliographystyle{apalike}
\bibliography{melodic.bib}  

\end{document}